\pdfoutput=1
\documentclass[journal]{IEEEtran}
\usepackage{draftwatermark}
\SetWatermarkText{Preprint}
\usepackage[latin9]{inputenc}
\usepackage{float}
\usepackage{bm}
\usepackage{amsmath}
\usepackage{amssymb}
\usepackage{graphicx}
\usepackage{esint}
\usepackage[T1]{fontenc}
\usepackage{url}
\usepackage{multirow}
\makeatletter
\usepackage{mathtools}
\DeclarePairedDelimiter\ceil{\lceil}{\rceil}
\DeclarePairedDelimiter\floor{\lfloor}{\rfloor}

\floatstyle{ruled}
\newfloat{algorithm}{tbp}{loa}
\providecommand{\algorithmname}{Algorithm}
\floatname{algorithm}{\protect\algorithmname}

\ifCLASSINFOpdf
\else
\fi
\usepackage[usenames,dvipsnames]{xcolor}\usepackage{pgfplots}\usepackage{amsfonts}\usepackage{epstopdf}\usepackage{bm}\usepackage{algorithmic}\usepackage{algorithm}
\usepackage{soul}
\usepackage{booktabs}
\floatstyle{ruled}
\newfloat{algorithm}{tbp}{loa}
\providecommand{\algorithmname}{Algorithm}
\floatname{algorithm}{\protect\algorithmname}
\newmuskip\pFqmuskip

\newcommand*{\pFq}[6][8]{%
  \begingroup 
  \pFqmuskip=#1mu\relax
  \mathchardef\normalcomma=\mathcode`,
  \mathcode`\,=\string"8000
  \begingroup\lccode`\~=`\,
  \lowercase{\endgroup\let~}\pFqcomma
  {}_{#2}F_{#3}%
  \endgroup
}
\newcommand{\pFqcomma}{{\normalcomma}\mskip\pFqmuskip}

\usepackage{cite}\usetikzlibrary{external}
  \tikzexternaldisable
\usetikzlibrary{plotmarks}
\pgfplotsset{compat = newest}

\definecolor{mygreen}{rgb}{0,0.5,0}
\definecolor{darkblue}{RGB}{0,0,150}
\makeatother

\begin{document}
\newcommand*{\xbar}[1]{%
   \hbox{%
     \vbox{%
       \hrule height 0.5pt 
       \kern0.25ex
       \hbox{%
         \kern-0.1em
         \ensuremath{#1}%
         \kern-0.1em
       }%
     }%
   }%
} 

\title{Bayesian Pitch Tracking Based on the Harmonic Model}


\author{Liming~Shi%
\thanks{This work was funded by the Danish Council for Independent Research,
grant ID: DFF 4184-00056.%
}%
\thanks{L. Shi, J. K. Nielsen, J. R. Jensen and M. G. Christensen are with
the Audio Analysis Lab, CREATE, Aalborg University, DK-9000 Aalborg,
Denmark, e-mail: \{ls, jrj, jkn,mgc\}@create.aau.dk%
},~\IEEEmembership{Student~Member,~IEEE,} Jesper~Kj\ae r~Nielsen,~\IEEEmembership{Member,~IEEE,}
Jesper~Rindom~Jensen,~\IEEEmembership{Member,~IEEE,} Max~A.~Little%
\thanks{M. A. Little is with the Engineering and Applied Science, Aston University
and Media Lab, Massachusetts Institute of Technology, e-mail: max.little@aston.ac.uk%
}, and~Mads~Gr\ae sb\o ll~Christensen,~\IEEEmembership{Senior~Member,~IEEE}
}

\maketitle
\begin{abstract}
Fundamental frequency  is one of the most important
characteristics of speech and audio signals. Harmonic model-based
 fundamental frequency estimators offer a higher estimation accuracy  and robustness against noise than the widely used autocorrelation-based methods. However, the traditional harmonic model-based estimators do not take the temporal smoothness of the fundamental frequency, the model order, and the voicing into account as they process each data segment independently. In this paper, a fully Bayesian fundamental frequency tracking algorithm based on the harmonic model and a first-order Markov process model is proposed. Smoothness priors are imposed on the fundamental frequencies, model orders, and voicing using first-order Markov process models.  Using these Markov models, fundamental frequency estimation and voicing detection errors can be reduced. Using the harmonic model, the proposed fundamental frequency tracker has an improved robustness to noise. An analytical form of the likelihood function, which can be  computed efficiently, is derived. Compared to the state-of-the-art neural network and non-parametric approaches,  the proposed  fundamental frequency tracking algorithm reduces the mean absolute errors and gross errors by 15\% and 20\% on the Keele pitch database and  36\% and 26\% on sustained
/a/ sounds from a database of Parkinson's disease voices under 0 dB white Gaussian noise. A MATLAB version of the proposed algorithm is made freely available for reproduction of the results\footnote{An implementation of the proposed algorithm using MATLAB may be found in \url{https://tinyurl.com/yxn4a543}}.
\end{abstract}

\begin{IEEEkeywords}
Fundamental frequency or pitch tracking,  harmonic model, Markov process, harmonic order, voiced-unvoiced
detection
\end{IEEEkeywords}
\IEEEpeerreviewmaketitle

\section{Introduction}

\IEEEPARstart{T}{he} problem of estimating the fundamental frequency or pitch information from noisy sound signals occurs in many applications, such as speech synthesis \cite{erro2014},
voice disorder detection \cite{tsanas2010}, and automatic speech recognition
\cite{ghahremani2014}. Fundamental frequency is a physical feature defined as the lowest frequency component of a periodic signal, while pitch is a perceptual feature, related to human listening \cite{gerhard2003pitch}. Our objective is to estimate fundamental frequency. But, following \cite{gonzalez2014pefac, Mads2008}, we do not distinguish between fundamental frequency and pitch and use them interchangeably. Pitch is usually estimated using a segment of sound signals (a.k.a., frame) with a fixed segment length (e.g., 15-35 ms for speech signals \cite{paliwal2010preference}). Numerous pitch estimation algorithms
have been proposed in the last fifty years, which can be categorized
as {}{unsupervised} and {}{supervised} approaches. Unsupervised
pitch estimation methods can be further categorized as {}{non-parametric}
and {}{parametric} approaches. Examples of non-parametric approaches
include the YIN \cite{de2002yin}, RAPT \cite{talkin1995robust},  SWIPE \cite{camacho2008sawtooth} and
PEFAC \cite{gonzalez2014pefac} methods. YIN and RAPT compute autocorrelation
functions from short frames of sound signals in the time domain. However, they are not robust against noise \cite{nielsen2017fast} and suffer from pitch octave errors
(that is, a rational multiple of the true pitch) \cite{ghahremani2014pitch}. To reduce the pitch octave errors, SWIPE uses the cross-correlation function against a sawtooth signal combined with the spectrum of the signal, and exploits only the first and prime harmonics of the signal. PEFAC estimates the pitch in the log-frequency
domain by convolving each frame's power spectrum with a filter that
sums the energy of the pitch harmonics. Dynamic programming is used to obtain a smooth estimate of the pitch track. Due to the filtering and built-in spectral normalization methods, PEFAC is claimed to work in high levels of noise.  However, a long frame length  (e.g., 90.5 ms in PEFAC by default)
is required to obtain good pitch estimation accuracy which is not
practical in many real-time applications.  More recently, a single frequency filtering approach based pitch estimation algorithm is proposed, which exploits the high SNR frequency component to overcome the effects of degradations in speech signal \cite{aneeja2017extraction}.

By contrast, parametric
methods (e.g., harmonic model-based pitch estimators \cite{quinn1991estimating, Mads2008,Sward2018})
have also been proposed for pitch estimation. Compared with non-parametric
approaches, harmonic model-based pitch estimators work with a short
frame length (e.g., 20 ms), and show higher robustness to additive noise, fewer
octave errors, and better time-frequency resolution \cite{ christensen07joint,6521410}. Recently,
a computationally efficient pitch estimator based on a harmonic model
has been proposed, which is referred to as the fast NLS \cite{nielsen2017fast}. However, one problem with most of the harmonic model based pitch estimators is that they do not take the temporal smoothness of the pitch, the harmonic order, and  voicing into account as they process each frame independently. As a result, outliers, due to octave errors
or voicing detection errors, occur. A sample-by-sample Kalman filtering-based pitch tracking algorithm using a time-varying
harmonic model is proposed in \cite{shi2017kalman} by assuming that
the pitch and weights follow first-order Markov chains.  A particle filtering-based pitch tracking algorithm based on the source-filter speech model combining with the harmonic modelling of input source is introduced in \cite{zhang2016fundamental}. However, the good performance of the algorithms in \cite{shi2017kalman} and \cite{zhang2016fundamental}
requires careful initializations. Moreover, it is difficult to integrate the time-varying
model order into these algorithms, see \cite{andrieu2003efficient} as an example of combing discrete and continuous state spaces. With either a known or
estimated model order, a maximum a posteriori (MAP) pitch estimator
based on the harmonic model has been developed to exploit the temporal
dynamics of the pitch \cite{tabrikian2004maximum}. The
model weights and observation noise variance are estimated by maximizing the maximum likelihood function (i.e., a frequentist perspective). Smooth pitch estimates are obtained,
and thus the pitch octave errors are reduced. An additional voicing
state is further considered in \cite{fisher2006generalized} for estimating
the pitch and obtaining the voiced-unvoiced decision jointly. However, the pitch tracking approach in \cite{tabrikian2004maximum}
and \cite{fisher2006generalized} has many drawbacks. First, the assumption
of a fixed harmonic order for multiple frames is not valid. In fact,
in audio signals, the harmonic order often changes from frame to frame \cite{norholm2016instantaneous}.
Second, matrix inversions are required to be stored for each
candidate pitch to reduce the computational complexity. Third, errors can be found in transition frames where the voicing
changes, because the past pitch information is not exploited when
an unvoiced frame occurs. Finally, it is well-known that estimating parameters from a frequentist's perspective leads to over-fitting \cite{bishop2006pattern}. 

More recently,
neural network based supervised pitch tracking algorithms were proposed
\cite{han2014neural,zhang2016pairwise,8461329}, which show robustness against noise. The method proposed in \cite{zhang2016pairwise} uses deep stacking network for joint speech separation and pitch estimation.  The CREPE \cite{8461329} discretises the pitch in logarithmic scale and uses a deep convolutional neural network to produce a pitch estimate. However, the unvoiced/silent state is not considered in the model. The maximum value of the output of the neural network is used as a heuristic estimate of the voicing probability. Moreover, to satisfy user's demand for different
frequency resolution or frame length, the whole system is required
to be retrained, which is usually time-consuming.

In this paper, we propose a fully {}{Bayesian} harmonic model-based
pitch tracking approach. By using the harmonic model, as opposed to non-parametric methods, improved robustness against background noise and octave errors can
be obtained. First-order Markov processes are used to capture the
temporal dynamics of pitch, harmonic order, and voicing.
By using information from previous frames, the rate of octave errors and
the voicing detection errors can be further reduced. Compared to \cite{tabrikian2004maximum} and \cite{fisher2006generalized},
we not only consider the temporal dynamics of pitch and voicing, but
also of the harmonic order, which enables us to detect if any pitch
is present, and estimate the pitch and harmonic order jointly and accurately. Moreover,
past information on pitch is exploited to improve robustness against
temporal voicing changes. Furthermore, by adopting a fully Bayesian
approach to model weights and observation noise variances, the overfitting can be
avoided. By assigning a proper transition pdf for the weights,
fast NLS \cite{nielsen2017fast} can be easily incorporated into the
proposed algorithm, leading to low computational and storage complexities. 

The rest of the paper is organized as follows. In Section \ref{sec:BayesTrack},
we briefly review general Bayesian tracking theory. In Section \ref{sec:hom}
and Section \ref{sec:sem}, we present the proposed harmonic observation
and state evolution models, respectively. In Section \ref{sec:pt},
the proposed pitch tracking algorithm is derived based on the harmonic
observation and state evolution models. In Section \ref{sec:prew},
we briefly review the prewhitening step for dealing with non-Gaussian noise. Simulation results are given
in Section \ref{sec:sim}, and the conclusions given in Section \ref{sec:con}.

Notation: Boldface symbols in lowercase and uppercase letters denote
column vectors and matrices, respectively.

\section{Bayesian tracking}

\label{sec:BayesTrack} In this section, we briefly review Bayesian
tracking in general, which forms the fundamental structure of the
proposed pitch tracking algorithm. Consider the problem of estimating
the state sequence $\{\mathbf{x}_{n}\},1\leq n\leq N$ from noisy
observations $\{\mathbf{y}_{n}\},1\leq n\leq N$, related by 
\begin{align}
\mathbf{y}_{n}=h(\mathbf{x}_{n},\mathbf{v}_{n}),\label{measure}
\end{align}
where $h(\cdot)$ denotes a mapping function between the state and
observation vectors, $\mathbf{v}_{n}$ denotes an i.i.d.
observation noise sequence, and $n$ denotes the time index. The state sequence follows a first-order
Markov process: 
\begin{align}
\mathbf{x}_{n}=f(\mathbf{x}_{n-1},\mathbf{m}_{n}),\label{transfer}
\end{align}
where $f(\cdot)$ denotes a mapping function between the current and
previous states, and $\mathbf{m}_{n}$ denotes an i.i.d.
state noise sequence. The  elements in the state vector $\mathbf{x}_{n}$ can either
be continuous or discrete. Assume that the posterior pdf $p(\mathbf{x}_{n-1}|\mathbf{Y}_{n-1})$
is available with the initial pdf being defined as $p(\mathbf{x}_{0})$,
where $\mathbf{Y}_{n-1}$ denotes a collection of observation vectors from the first
observation vector up to the $({n-1})^{\mathrm{th}}$ observation vector, i.e.,
\begin{align}
\mathbf{Y}_{n-1}=[\mathbf{y}_{1},\cdots,\mathbf{y}_{n-1}].  \nonumber 
\end{align}
The
objective of Bayesian tracking is to obtain a posterior distribution
over the state vector $\mathbf{x}_{n}$ based on the current and previous
observations recursively, i.e., $p(\mathbf{x}_{n}|\mathbf{Y}_{n})$.
The posterior $p(\mathbf{x}_{n}|\mathbf{Y}_{n})$ can be obtained
in two stages: predict and update. 

In the prediction stage, we obtain the prediction pdf $p(\mathbf{x}_{n}|\mathbf{Y}_{n-1})$
by using the transition pdf $p(\mathbf{x}_{n}|\mathbf{x}_{n-1})$
from \eqref{transfer}, i.e., 
\begin{align}
 & p(\mathbf{x}_{n}|\mathbf{Y}_{n-1})  \nonumber \\
= & \int p(\mathbf{x}_{n}|\mathbf{x}_{n-1})p(\mathbf{x}_{n-1}|\mathbf{Y}_{n-1})d\mathbf{x}_{n-1},  \ \ \  2\leq n \leq N,  \nonumber \\
 &p(\mathbf{x}_{1})=  \int p(\mathbf{x}_{1}|\mathbf{x}_{0})p(\mathbf{x}_{0})d\mathbf{x}_{0},\ \ \ n=1,
\label{prediction}
\end{align}
which is known as the Chapman-Kolmogorov equation. Note that if the elements in $\mathbf{x}_{n}$
are all discrete variables, the integration operator should be replaced with the
summation operator.

In the update stage, combining \eqref{measure} and the prediction
pdf from the prediction stage, Bayes' rule can be applied to obtain
the posterior, i.e., 
\begin{align}
p(\mathbf{x}_{n}|\mathbf{Y}_{n})&=\frac{p(\mathbf{y}_{n}|\mathbf{x}_{n},\mathbf{Y}_{n-1})p(\mathbf{x}_{n}|\mathbf{Y}_{n-1})}{p(\mathbf{y}_{n}|\mathbf{Y}_{n-1})}, \ \ \  2\leq n \leq N,   \nonumber \\
&p(\mathbf{x}_{1}|\mathbf{Y}_{1})=\frac{p(\mathbf{y}_{1}|\mathbf{x}_{1})p(\mathbf{x}_{1})}{p(\mathbf{y}_{1})}, \ \ \ n=1,
\label{update}
\end{align}
where $p(\mathbf{y}_{n}|\mathbf{x}_{n},\mathbf{Y}_{n-1})$  and  $p(\mathbf{y}_{1}|\mathbf{x}_{1})$ are the
likelihood functions and $p(\mathbf{y}_{n}|\mathbf{Y}_{n-1})$ and $p(\mathbf{y}_{1})$ are
the normalization factors, respectively. Closed form solutions can be obtained for
\eqref{prediction} and \eqref{update} in at least two cases. In the first
case, when both $\mathbf{v}_{n}$ and $\mathbf{m}_{n}$ are drawn
from Gaussian distributions with known parameters, and both $h(\mathbf{x}_{n},\mathbf{v}_{n})$
and $f(\mathbf{x}_{n-1},\mathbf{m}_{n})$ are linear functions over
the variables, \eqref{prediction} and \eqref{update} reduce to the
well-known Kalman-filter\cite{bishop2006pattern}. In the
second case, when the state space is discrete and has a limited number
of states, \eqref{prediction} and \eqref{update} reduce to the forward
step of the forward-backward algorithm for hidden Markov model (HMM)
inference \cite{bishop2006pattern}. In other cases, the inference
of the posterior $p(\mathbf{x}_{n}|\mathbf{Y}_{n})$ can be approximated
using Monte Carlo approaches, such as particle filtering \cite{arulampalam2002tutorial}.
Next, we define the mapping function $h(\cdot)$ and formulate the
observation equation \eqref{measure} based on the harmonic model
in Section \ref{sec:hom}, and then explain the state evolution
model \eqref{transfer} for the proposed pitch tracking algorithm
in Section \ref{sec:sem}.

\section{Harmonic observation model}
\label{sec:hom}
\subsection{The harmonic observation model}

Consider the general signal observation model given by
\begin{align}
\mathbf{y}_{n}=\mathbf{s}_{n}+\mathbf{v}_{n},\label{observation_eq}
\end{align}
where the observation vector $\mathbf{y}_{n}$ is a collection of $M$ samples from the $n^{\mathrm{th}}$ frame defined as
$$\mathbf{y}_{n}=[y_{n,1},\cdots,y_{n, M}]^{T},$$
the clean signal vector $\mathbf{s}_{n}$ and noise vector $\mathbf{v}_{n}$ are defined similarly to $\mathbf{y}_{n}$, $M$ is the frame length in samples and $n$ is the frame index. We assume that $\mathbf{v}_{n}$
is a multivariate white noise processes with zero mean and diagonal covariance matrix
$\sigma_{n}^{2}\mathbf{I}$, $\sigma_{n}^{2}$ is the noise variance, $\mathbf{I}$ is the identity matrix. When voiced speech or music is present,
we assume that the pitch, model weights and model order are constant
over a short frame (typically 15 to 35 ms for speech signals and longer for
music signals) and $s_{n,m}$ (i.e., the $m^{\text{th}}$ element of $\mathbf{s}_{n}$)
follows the harmonic model, i.e., 
\begin{align}
\text{H}_{1}:s_{n,m} & =\sum_{k=1}^{K_{n}} \left[\alpha_{k,n}\text{cos}(k\omega_{n}m)+\beta_{k,n}\text{sin}(k\omega_{n}m)\right],\label{Harmonic_eq}
\end{align}
where $\alpha_{k,n}$ and $\beta_{k,n}$ are the linear weights of the
$k^{\mathrm{th}}$ harmonic, $\omega_{n}={2\pi f_{n}}/{f_{\text{s}}}$ is
the normalized digital radian frequency, $f_{\text{s}}$ is the sampling
rate, and $K_{n}$ is the number
of harmonics. When voiced speech/music is absent (unvoiced
or silent), a null model is used, i.e., 
\begin{align}
\text{H}_{0}:\mathbf{y}_{n} & =\mathbf{v}_{n}.\label{null_matrix_eq}
\end{align}
Note that, based on the source-filtering model of speech generation,
the unvoiced speech can be modelled as a coloured Gaussian process \cite{makhoul1975linear}.
The observation noise in practice may have non-stationary and non-Gaussian properties, such as babble noise. However, we can deal with this by prewhitening the
observation signals \cite{norholm2016instantaneous}, which will be described in Section \ref{sec:prew}. Writing \eqref{Harmonic_eq}
in matrix form and combining \eqref{observation_eq} and \eqref{Harmonic_eq}
yields 
\begin{align}
\text{H}_{1}:\mathbf{y}_{n} & =\mathbf{Z}(\omega_{n},K_{n})\mathbf{a}_{K_{n}}+\mathbf{v}_{n},\label{harmonic_matrix_eq}
\end{align}
where 
\begin{align}
\mathbf{Z}(\omega_{0},K_{n})&=[\mathbf{c}(\omega_{n}),\cdots,\mathbf{c}(K_{n}\omega_{n}),\mathbf{d}(\omega_{n}),\cdots,\mathbf{d}(K_{n}\omega_{n})],\nonumber \\
 \mathbf{c}(\omega_{n})&=[\text{cos}(\omega_{n}1),\cdots,\text{cos}(\omega_{n}M)]^{T},\nonumber \\
\mathbf{d}(\omega_{n})&=[\text{sin}(\omega_{n}1),\cdots,\text{sin}(\omega_{n}M)]^{T},\nonumber \\
\mathbf{a}_{K_{n}}&=[\alpha_{1,n},\cdots,\alpha_{K_{n},n},\beta_{1,n},\cdots,\beta_{K_{n},n}]^{T}. \nonumber
\end{align}
We can further write \eqref{null_matrix_eq} and \eqref{harmonic_matrix_eq}
together by introducing a binary voicing indicator variable $u_{n}$,
i.e., 
\begin{align}
\mathbf{y}_{n} & ={u_{n}}\mathbf{Z}(\omega_{n},K_{n})\mathbf{a}_{K_{n}}+\mathbf{v}_{n},\label{finalmodel}
\end{align}
where $u_{n}\in\{0,1\}$. When $u_{n}=0$ and $u_{n}=1$, \eqref{finalmodel}
reduces to the unvoiced and voiced models \eqref{null_matrix_eq}
and \eqref{harmonic_matrix_eq}, respectively.

We collect all the unknown variables into the state vector $\mathbf{x}_{n}=[\mathbf{a}_{K_{n}},\sigma_{n}^{2},\omega_{n},K_{n},u_{n}]^{T}$.
Comparing \eqref{finalmodel} and \eqref{measure}, we can conclude
that the mapping function $h(\cdot)$ is a nonlinear function w.r.t.
the state vector $\mathbf{x}_{n}$. Moreover, the state vector $\mathbf{x}_{n}$
contains continuous variables $\mathbf{a}_{K_{n}}$, $\sigma_{n}^{2}$,
$\omega_{n}$ and discrete variables $K_{n}$ and $u_{n}$. However,
due to the non-linear characteristics of \eqref{finalmodel} w.r.t.
$\omega_{n}$, uniform discretisation over the pitch $\omega_{n}$
is commonly used \cite{nielsen2017fast}. An off-grid estimate of $\omega_{n}$ can be obtained
by pitch refinement algorithms, such as gradient descent \cite{christensen2008multi}.
Our target is to obtain estimates of the fundamental frequency $\omega_{n}$,
the harmonic order $K_{n}$, and the voicing indicator $u_{n}$, that
is a subset of $\mathbf{x}_{n}$ defined as $\ddot{\mathbf{x}}_{n}=[\omega_{n},K_{n},u_{n}]^{T}$,
from the noisy observation $\mathbf{y}_{n}$.

\section{The state evolution model}

\label{sec:sem} In this section, we derive the state evolution model
\eqref{transfer} or more generally the transition probability density/mass
function (pdf/pmf) $p(\mathbf{x}_{n}|\mathbf{x}_{n-1},\mathbf{Y}_{n-1})$ for continuous/discrete states of the proposed model.  
Following the  fast NLS pitch estimation approach \cite{nielsen2017fast}, we uniformly discretize
the pitch $\omega_{n}\in\{\omega^{f},1\leq f\leq F\}$ over the range
$[\omega_{\min},\omega_{\max}]$,  where $\omega_{\min}$ and $\omega_{\max}$ denote the lowest and highest pitches in the searching space, respectively. Prior information can be used to set $\omega_{\min}$ and $\omega_{\max}$. For example, pitch is usually between 70 to 400 Hz for speech signals. The grid size is set to 
\begin{align}
\floor*{F\frac{\omega_{\max}}{2\pi}}-\ceil*{F\frac{\omega_{\min}}{2\pi}}+1,  \nonumber 
\end{align}
where $F$ denotes the DFT size for computing the likelihood function (see Section \ref{sec:pt} and \cite{nielsen2017fast} for further details), $\floor*{\cdot}$ and $\ceil*{\cdot}$ denote the flooring and ceiling operators, respectively.  It is also shown that the optimal choice of $F$ depends on the frame length and the harmonic order \cite{nielsen2017fast}. However, for simplicity and fast implementation, in this paper, we set $F=2^{14}$. The state space for the discrete
variables can be expressed as $\{\mathcal{M}(n):[\omega_{n}=\omega^{f},K_{n}=k,u_{n}=1]^{T},1\leq f\leq F,1\leq k\leq K^{\max}\}\cup\{\mathcal{M}_{0}(n):u_{n}=0\}$.
The prediction pdf $p(\mathbf{x}_{n}|\mathbf{Y}_{n-1})$ defined in
\eqref{prediction} can be factorized as 
\begin{align}
p(\mathbf{x}_{n}|\mathbf{Y}_{n-1})= & p(\mathbf{a}_{K_{n}}|\sigma_{n}^{2},\ddot{\mathbf{x}}_{n},\mathbf{Y}_{n-1})\times\nonumber \\
 & p(\sigma_{n}^{2}|\ddot{\mathbf{x}}_{n},\mathbf{Y}_{n-1})p(\ddot{\mathbf{x}}_{n}|\mathbf{Y}_{n-1}).
\end{align}
We first explain the transition pdfs for the continuous variables
$\sigma_{n}^{2}$ and $\mathbf{a}_{K_{n}}$, and then discuss the
transition pmfs for the discrete variables $\omega_{n}$, $K_{n}$
and $u_{n}$.  The selection of a state evolution model is a trade-off between being physically accurate and ending up with a practical solution.

\subsection{Transition pdfs for the noise variance and weights}

To obtain the prediction pdf for the noise variance $p(\sigma_{n}^{2}|\ddot{\mathbf{x}}_{n},\mathbf{Y}_{n-1})$,
the transition pdf for the noise variance $p(\sigma_{n}^{2}|\sigma_{n-1}^{2},\ddot{\mathbf{x}}_{n},\mathbf{Y}_{n-1})$
should be defined. A reasonable assumption for the noise variance
is that it changes slowly from frame to frame. For example, the unknown parameter $\sigma_{n}^{2}$ can be assumed to evolve according to
an inverse Gamma distribution \cite{fevotte2009nonnegative}, i.e. 
\begin{align}
p(\sigma_{n}^{2}|\sigma_{n-1}^{2})=\mathcal{IG}(\sigma_{n}^{2}|c,d\sigma_{n-1}^{2}).\label{sigmaprior2}
\end{align}
where $\mathcal{IG}(x| \alpha,\beta)= \frac{\beta^\alpha}{\Gamma(\alpha)}x^{-\alpha-1}\exp(-\frac{\beta}{x})$ and $\Gamma(\cdot)$ denotes the gamma function.
With this transition pdf, an analytical form of the posterior distribution
on $\mathbf{x}_{n}$ cannot be derived. A sequential Monte Carlo approach
can be used to approximate the posterior numerically \cite{cappe2007overview}.
However, the major drawback of any Monte Carlo filtering strategy
is that sampling in high-dimensional spaces can be inefficient \cite{fong2002monte}.
A Rao-blackwellized particle filtering approach \cite{doucet2000rao},
which marginalises out some of the variables for statistical variance
reduction, can be used to deal with this problem. However, we do not pursue this approach any further in this paper, and leave it for future work. Instead, for simplicity, we assume independence
between $\sigma_{n}^{2}$ and $\sigma_{n-1}^{2}$, and use the Jeffery's
prior, i.e., 
\begin{align}
p(\sigma_{n}^{2}|\sigma_{n-1}^{2},\ddot{\mathbf{x}}_{n},\mathbf{Y}_{n-1})\propto1/\sigma_{n}^{2},\sigma_{n}^{2}>0.\label{sigmaprior}
\end{align}
As can be seen, the Jeffery's prior \eqref{sigmaprior} is a limiting case of \eqref{sigmaprior2} with $c\to0$ and $d\to0$.

Similarly, we define the transition pdf for the weights as $p(\mathbf{a}_{K_{n}}|\mathbf{a}_{K_{n-1}},\sigma_{n}^{2},\ddot{\mathbf{x}}_{n},\mathbf{Y}_{n-1})$.
Imposing smoothness dependency on the weight time evolution can
reduce pitch octave errors\cite{adalbjornsson2015multi}. However,
in order to use the fast algorithm \cite{nielsen2017fast}, we assume
that the model weights between consecutive frames are conditionally independent given previous observations and the rest of unknown variables.
Following \cite{nielsen2014bayesian}, we use the hierarchical prior
\begin{align}
 & p(\mathbf{a}_{K_{n}}|\mathbf{a}_{K_{n-1}},\sigma_{n}^{2},\ddot{\mathbf{x}}_{n},\mathbf{Y}_{n-1},g_{n})\nonumber \\
= & \mathcal{N}(\mathbf{a}_{K_{n}}|0,g_{n}\sigma_{n}^{2}\left[(\mathbf{Z}(\omega_{n},K_{n})^{T}\mathbf{Z}(\omega_{n},K_{n})\right]^{-1}),\label{gprior}\\
&p(g_{n}|\delta)  =\frac{\delta-2}{2}(1+g_{n})^{-\delta/2},g>0,\label{ghyperprior}
\end{align}
where $\mathcal{N}(\mathbf{x}|\bm{\mu},\bm{\Sigma})$ denotes that
the vector $\mathbf{x}$ has the multivariate normal distribution
with mean $\bm{\mu}$ and covariance matrix $\bm{\Sigma}$. The prior
distribution for the weights \eqref{gprior} is known as Zellner's
g-prior \cite{zellner1986assessing}. As can be seen from \eqref{gprior}, given $\omega_{n}$ and $K_{n}$, the prior covariance matrix is a scaled version of the Fisher information matrix. With Zellner's g-prior, a closed-form calculation of the marginal likelihood can be obtained \cite{liang2008mixtures}. Moreover, the fast algorithm in \cite{nielsen2017fast} for computing the marginal likelihood can be applied (see Section \ref{sec:pt} for detail).

\begin{figure}[t]
\includegraphics[scale=0.9]{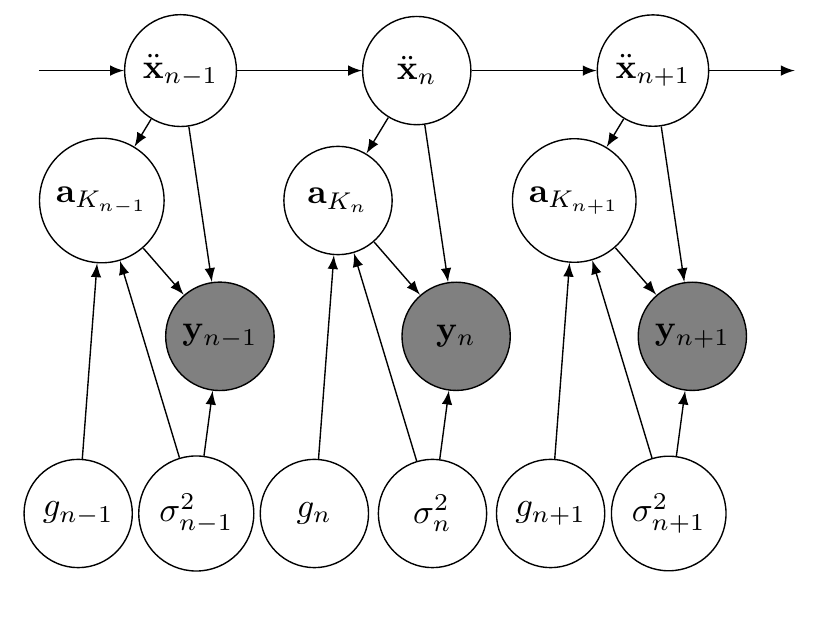}
\caption{A graphical model of the proposed method with shaded nodes indicating
observed variables.}

\label{fig:states} 
\end{figure}

The graphical model for the proposed method is shown in Fig.\ \ref{fig:states}.
Note that, instead of obtaining point estimates of the noise variance
and weight parameters using maximum likelihood \cite{tabrikian2004maximum},
a Bayesian approach is used to represent the full uncertainty over
these parameters. 

\subsection{Transition pmfs for $\omega_{n},K_{n}$ and $u_{n}$}

\begin{figure}[t]
\includegraphics[scale=0.9]{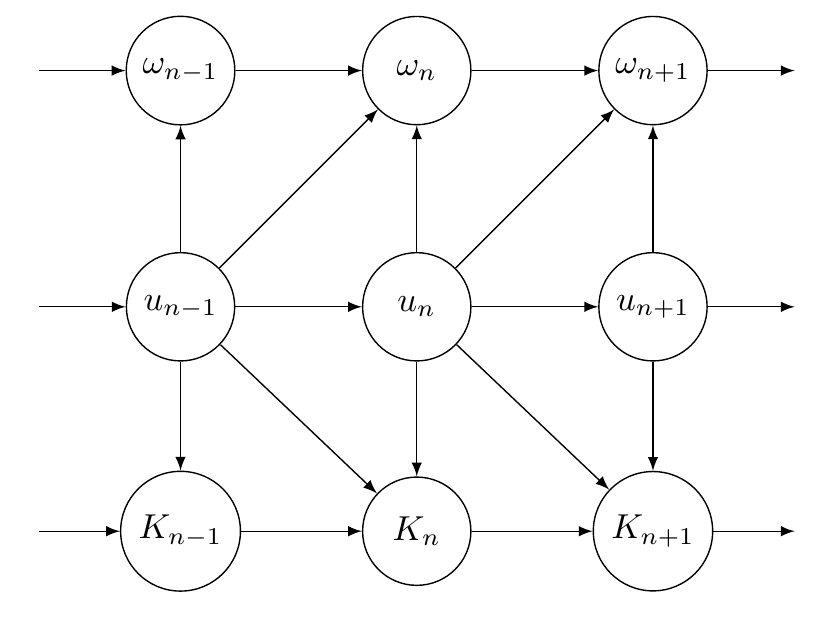}
\caption{A graphical model specifying conditionally independence relations for
the discrete variables.}

\label{fig:royaaa} 
\end{figure}

In \cite{tabrikian2004maximum}, to reduce octave errors, a first-order Markov model is used for the pitch evolution
provided that the harmonic order is fixed and known/estimated for
multiple frames. Another voicing evolution model is further considered
in \cite{fisher2006generalized} by imposing the so-called \textquotedbl{}hang-over\textquotedbl{}
scheme \cite{sohn1999statistical}. Although in some cases, the harmonic
order may not be of interest, it is still necessary to estimate it to obtain
correct pitch estimates \cite{christensen2007joint}. In fact, considering the temporal dynamics of the model order helps reducing the octave errors, which will be verified by the simulation results. Moreover, using priors for the model order is also necessary for model comparison \cite{nielsen2014bayesian}. In this paper,
we propose to track the pitch $\omega_{n}$, the harmonic order $K_{n}$
and the voicing indicator $u_{n}$ jointly. More specifically, we
impose smoothness constraints on $\omega_{n}$ and $K_{n}$, and hang-over
on voicing state using first-order Markov processes. The transition
probability for the $n^{\mathrm{th}}$ frame to be voiced with pitch
$\omega_{n}$ and harmonic order $K_{n}$ when the previous frame
is also voiced with $\omega_{n-1}$ and $K_{n-1}$ can be expressed
as 
\begin{align}\label{totaltransfer}
 & p(\mathcal{M}(n)|\mathcal{M}(n-1))\nonumber \\
= & p(\omega_{n},K_{n}|\omega_{n-1},K_{n-1},u_{n-1}=1,u_{n}=1)\times\nonumber \\
 & p(u_{n}=1|u_{n-1}=1).
\end{align}
We assume that the pitch $\omega_{n}$ and harmonic
order $K_{n}$ evolve according to their own, independent dynamics given $u_n=1$ and $u_{n-1}=1$,
i.e., 
\begin{align} \label{assump1}
 & p(\omega_{n},K_{n}|\omega_{n-1},K_{n-1},u_{n}=1,u_{n-1}=1)\nonumber \\
= & p(\omega_{n}|\omega_{n-1},u_{n}=1,u_{n-1}=1)\times\nonumber \\
 & p(K_{n}|K_{n-1},u_{n}=1,u_{n-1}=1),
\end{align}
which means when both time frame $n-1$ and $n$ are voiced, the pitch
and harmonic order only depend on their previous states. In fact, this assumption is only true when the product of the maximum allowed harmonic order and the pitch is less than half of the sampling frequency. However, by using a Bayesian approach, a model with a larger harmonic order is more penalized than with a smaller one.  Even if a large value is used for the maximum allowed harmonic order in the proposed approach, the posterior model probability with a large harmonic order can be small \cite{stoica2004model}. In \cite{yoshii2012nonparametric}, an infinite number of harmonics is used, and the non-parametric prior distribution is used to penalize the models with large harmonic orders. By assuming the pitch and harmonic order are conditionally independent given $u_n=1$ and $u_{n-1}=1$, the Bayesian inference of the model posterior, shown in Section \ref{sec:pt},  can be simplified. The transition
probability for the $n^{\mathrm{th}}$ frame to be voiced with pitch
$\omega_{n}$ and harmonic order $K_{n}$ when the previous frame
is unvoiced/silent can be expressed as 
\begin{align}\label{unvoicedtovoiced}
 & p(\mathcal{M}(n)|\mathcal{M}_{0}(n-1))\nonumber \\
= & p(\omega_{n},K_{n}|u_{n}=1,u_{n-1}=0)p(u_{n}=1|u_{n-1}=0).
\end{align}
The priors from an unvoiced frame to a voiced frame $p(\omega_{n},K_{n}|u_{n}=1,u_{n-1}=0)$
are set to $p(\omega_{m},K_{m}|\mathbf{Y}_{m},u_{m}=1)$, which can
be calculated as 
\begin{align}
p(\omega_{m},K_{m}|\mathbf{Y}_{m},u_{m}=1)=\frac{p(\omega_{m},K_{m},u_{m}=1|\mathbf{Y}_{m})}{1-p(u_{m}=0|\mathbf{Y}_{m})},\label{pastinfo}
\end{align}
where $m$ is the closest frame index to $n$ that satisfies the constraint
$p(u_{m}=0|\mathbf{Y}_{m})<0.5$ ($m^{\mathrm{th}}$ frame is voiced).
In fact, if the previous frame is not voiced, we exploit the information
from the latest frame that is voiced as the prior for the pitches
and harmonic orders. The motivation for this choice is that the pitch
and harmonic order usually do not change abruptly after a short segment of
unvoiced/silent frames. Using the past information as the prior, robustness
against the voicing state changes can be improved. The graphical model
for the evolution of $\ddot{\mathbf{x}}(n)$ is shown in Fig.\ \ref{fig:royaaa}.
Assuming the Markov processes are time-invariant, we can express the
transition matrices for the pitch, harmonic order and voicing
as $\mathbf{A}^{\omega}$, $\mathbf{A}^{K}$ and $\mathbf{A}^{u}$,
respectively. 

\section{Pitch tracking}
\label{sec:pt}
 In this section, a joint pitch and harmonic order
tracking, and voicing detection algorithm is derived based on the
Bayesian tracking formulas \eqref{prediction} and \eqref{update}.
First, note that, by assuming that $\sigma_{n}^{2}$ and $\sigma_{n-1}^{2}$
are conditionally independent given $\ddot{\mathbf{x}}_{n}$ and $\mathbf{Y}_{n-1}$, and $\mathbf{a}_{K_{n}}$ and $\mathbf{a}_{K_{n-1}}$
are conditionally independent given $\sigma_{n}^{2}$, $\ddot{\mathbf{x}}_{n}$ and $\mathbf{Y}_{n-1}$, the prediction pdfs are equal to the
transition pdfs, i.e., 
\begin{align}
 & p(\sigma_{n}^{2}|\ddot{\mathbf{x}}_{n},\mathbf{Y}_{n-1})=p(\sigma_{n}^{2}|\sigma_{n-1}^{2},\ddot{\mathbf{x}}_{n},\mathbf{Y}_{n-1}),\label{psigma}\\
 & p(\mathbf{a}_{K_{n}}|\sigma_{n}^{2},\ddot{\mathbf{x}}_{n},\mathbf{Y}_{n-1})\nonumber \\
= & \int p(\mathbf{a}_{K_{n}}|\mathbf{a}_{K_{n-1}},\sigma_{n}^{2},\ddot{\mathbf{x}}_{n},\mathbf{Y}_{n-1},g_{n})p(g_{n};\delta)dg_{n}.\label{pweights}
\end{align}
Based on \eqref{prediction}, prediction pmfs for discrete variables
$p(\ddot{\mathbf{x}}(n)|\mathbf{Y}_{n-1})$ can be expressed as 
\begin{align} \label{voicingpredi}
 & p(\mathcal{M}(n)|\mathbf{Y}_{n-1})\nonumber \\
= & \sum_{\mathcal{M}(n-1)}p(\mathcal{M}(n)|\mathcal{M}(n-1))p(\mathcal{M}(n-1)|\mathbf{Y}_{n-1})+\nonumber \\
 & p(\mathcal{M}(n)|\mathcal{M}_{0}(n-1))p(\mathcal{M}_{0}(n-1)|\mathbf{Y}_{n-1}),
\end{align}
\begin{align} \label{unvoicingpredi}
 & p(\mathcal{M}_{0}(n)|\mathbf{Y}_{n-1})\nonumber \\
= & \sum_{h=0}^{1}p(u_{n}=0|u_{n-1}=h)p(u_{n-1}=h|\mathbf{Y}_{n-1})\nonumber \\
= & p(u_{n}=0|u_{n-1}=0)p(\mathcal{M}_{0}(n-1)|\mathbf{Y}_{n-1})+\nonumber \\
 & p(u_{n}=0|u_{n-1}=1)(1-p(\mathcal{M}_{0}(n-1)|\mathbf{Y}_{n-1})).
 \end{align}
With the prediction pdfs and pmfs in hand, we can obtain the update
equation based on \eqref{update}. In order to obtain the posteriors
for the pitch, harmonic order and voicing indicators, the weights
and noise variance can be integrated out from the update equation,
i.e., 
\begin{align} \label{updatefinal}
 & p(\ddot{\mathbf{x}}_{n}|\mathbf{Y}_{n})\nonumber \\
\propto & \int p(\mathbf{y}_{n}|\mathbf{x}_{n},\mathbf{Y}_{n-1})p(\mathbf{x}_{n}|\mathbf{Y}_{n-1})d\mathbf{a}_{K_{n}}d\sigma_{n}^{2}\nonumber \\
= & p(\mathbf{y}_{n}|\ddot{\mathbf{x}}_{n},\mathbf{Y}_{n-1})p(\ddot{\mathbf{x}}_{n}|\mathbf{Y}_{n-1}),
\end{align}
where $p(\mathbf{y}_{n}|\ddot{\mathbf{x}}_{n},\mathbf{Y}_{n-1})$
denotes a marginal likelihood, defined as 
\begin{align}\label{marginallikelihood}
p(\mathbf{y}_{n}|\ddot{\mathbf{x}}_{n},\mathbf{Y}_{n-1})= & \int p(\mathbf{y}_{n}|\mathbf{x}_{n})p(\mathbf{a}_{K_{n}}|\sigma_{n}^{2},\ddot{\mathbf{x}}_{n},\mathbf{Y}_{n-1})\times\nonumber \\
 & p(\sigma_{n}^{2}|\ddot{\mathbf{x}}_{n},\mathbf{Y}_{n-1})p(g_{n};\delta)d\mathbf{a}_{K_{n}}d\sigma_{n}^{2}dg_{n}.
\end{align}
Using \eqref{finalmodel}, \eqref{sigmaprior}, \eqref{gprior}, \eqref{ghyperprior},
\eqref{psigma} and \eqref{pweights}, a closed-form marginal likelihood
can be obtained, i.e., 
\begin{align}\label{likelihoodexp}
 & p(\mathbf{y}_{n}|\ddot{\mathbf{x}}_{n},\mathbf{Y}_{n-1})\nonumber \\
= & \left[\frac{(\delta-2)}{2K_{n}+\delta-2}{}_{2}F_{1}\left[\frac{M}{2},1;\frac{2K_{n}+\delta}{2};R^{2}(\omega_{n},K_{n})\right]\right]^{u_{n}}\times\nonumber \\
 & m_{M}(\mathbf{y}_{n}),
\end{align}
where 
\begin{align}\label{maginallike}
&m_{M}(\mathbf{y}_{n})=  \frac{\Gamma(\frac{M}{2})}{(\pi||\mathbf{y}_{n}||_{2}^{2})^{\frac{M}{2}}},\\
&R^{2}(\omega_{n},K_{n})=  \frac{\mathbf{y}_{n}^{T}\mathbf{Z}(\omega_{n},K_{n})\hat{\mathbf{a}}_{K_{n}}}{\mathbf{y}_{n}^{T}\mathbf{y}_{n}}, \\
& \hat{\mathbf{a}}_{K_{n}}=(\mathbf{Z}(\omega_{n},K_{n})^{T}\mathbf{Z}(\omega_{n},K_{n}))^{-1}\mathbf{Z}(\omega_{n},K_{n})\mathbf{y}_{n},
\end{align}
$m_{M}(\mathbf{y}_{n})$ denotes the null model likelihood (i.e.,
$p(\mathbf{y}_{n}|u_{n}=0)$) and $_{2}F_{1}$ denotes the Gaussian
hypergeometric function \cite{gradshteyn2014table}. To compute $R^{2}(\omega_{n},K_{n})$ for all the candidate pitches and harmonic orders, the fast algorithm \cite{nielsen2017fast} can be applied. Moreover, from a computational
point of view, a Laplace approximation of \eqref{marginallikelihood} can be
derived as an alternative instead of marginalizing w.r.t. $g_{n}$
analytically \cite{nielsen2014bayesian}. Note that, for the discrete
vector $\ddot{\mathbf{x}}_{n}$, it should satisfy the normalisation
constraint, 
\begin{align}\label{sumtoone}
1= & \sum_{\ddot{\mathbf{x}}_{n}}p(\ddot{\mathbf{x}}_{n}|\mathbf{Y}_{n})\nonumber \\
= & p(\mathcal{M}_{0}(n)|\mathbf{Y}_{n})+\sum_{\mathcal{M}(n)}p(\mathcal{M}(n)|\mathbf{Y}_{n}).
\end{align}

Finally, estimates of the pitch and harmonic order and the voiced/unvoiced
state can be jointly obtained using the maximum a posterior (MAP)
estimator. More specifically, the $n^{\mathrm{th}}$ frame is labeled
as voiced if $p(u_{n}=0|\mathbf{Y}_{n})<0.5$, and the pitch and harmonic
order are obtained as 
\begin{align}
(\hat{\omega}_{n},\hat{K}_{n})=\max_{\omega_{n},K_{n}}p(\omega_{n},K_{n},u_{n}=1|\mathbf{Y}_{n}).\label{finalest}
\end{align}
The proposed Bayesian pitch tracking algorithm is shown in Algorithm
\ref{alg:pitchtracking}. To make inferences, we need to specify
the transition matrices for the pitch $p(\omega_{n}|\omega_{n-1},u_{n}=1,u_{n-1}=1)$,
the harmonic order $p(K_{n}|K_{n-1},u_{n}=1,u_{n-1}=1)$ and $p(u_{n}|u_{n-1})$. Following \cite{tabrikian2004maximum}, we set $p(\omega_{n}|\omega_{n-1},u_{n}=1,u_{n-1}=1)=\mathcal{N}(\omega_{n}|\omega_{n-1},\sigma_{\omega}^{2})$.
The transition probability for the model order is chosen as $p(K_{n}|K_{n-1},u_{n}=1,u_{n-1}=1)=\mathcal{N}(K_{n}|K_{n-1},\sigma_{K}^{2})$.
Smaller $\sigma_{\omega}^{2}$ and $\sigma_{K}^{2}$ lead to smoother
estimates of the pitch and harmonic order while larger values make
the inference less dependent on the previous estimates. The matrix
$\mathbf{A}^{u}$ is controlled by $p(u_{n}=1|u_{n-1}=0)$ and $p(u_{n}=0|u_{n-1}=1)$.
In order to reduce the false negative (wrongly classified as unvoiced when a frame is voiced) rate, we set $p(u_{n}=1|u_{n-1}=0)=0.4$,
$p(u_{n}=0|u_{n-1}=1)=0.3$, respectively, that is, the transition
probability from unvoiced to voiced is higher than from voiced to
unvoiced. Note that each row of $\mathbf{A}^{\omega}$, $\mathbf{A}^{K}$, 
and $\mathbf{A}^{u}$ is normalised to ensure they are proper pmfs.
By setting $\sigma_{\omega}^{2}\to\infty$, $\sigma_{K}^{2}\to\infty$,
$p(u_{n}=1|u_{n-1}=0)=0.5$ and $p(u_{n}=0|u_{n-1}=1)=0.5$, the proposed
algorithm reduces to the fast NLS algorithm \cite{nielsen2017fast}.

\begin{algorithm}[t]
	\caption{The proposed Bayesian pitch tracking}
	\label{alg:pitchtracking}
	\begin{algorithmic} [1]
		\STATE Initiate the harmonic order $K^{\max}$, transition matrices $\mathbf{A}^\omega$, $\mathbf{A}^K$ and $\mathbf{A}^u$, and the initial probability $p(u_0|\mathbf{y}_0)$ and $p(\omega_0, K_0, u_0=1|\mathbf{y}_0)$ satisfying the constraint $p(u_0=0|\mathbf{y}_0)+\sum_{\omega_0, K_0}p(\omega_0, K_0, u_0=1|\mathbf{y}_0)=1$
		\FOR   {$n = 1,2, \cdots$}
		\STATE \textit{Prediction step: }
		\STATE Obtain $p(\mathcal{M}(n)|\mathbf{Y}_{n-1})$ based on \eqref{voicingpredi}, \eqref{totaltransfer} and \eqref{unvoicedtovoiced}.
		\STATE Obtain $p(\mathcal{M}_0(n)|\mathbf{Y}_{n-1}) $ based on \eqref{unvoicingpredi}.
		\STATE \textit{Update step: }		
		\STATE Calculate $p(\mathbf{y}_n| \ddot{\mathbf{x}}_n,\mathbf{Y}_{n-1})$ using the fast weight estimation algorithm \cite{nielsen2017fast} and \eqref{likelihoodexp}.
		\STATE Calculate the unnormalised posteriors $p(\mathcal{M}(n)|\mathbf{Y}_{n})$ and  $p(\mathcal{M}_0(n)|\mathbf{Y}_{n})$ based on \eqref{updatefinal}.
		\STATE Normalise the posteriors based on the constraint \eqref{sumtoone}.
		\STATE \textit{MAP estimation: }	
		\IF {$p(\mathcal{M}_0(n)|\mathbf{Y}_{n})>0.5$}
		\STATE The $n^{th}$ frame is labeled as unvoiced/silent.
		\ELSE
		\STATE The $n^{th}$ frame is labeled as voiced.
		\STATE Estimating $\hat{\omega}_n$ and $\hat{K}_n$ based on \eqref{finalest}.
		\STATE Update $p(\omega_{m}, K_{m}| \mathbf{Y}_{m},u_m=1)$ based on \eqref{pastinfo}.
		\ENDIF		
	         \ENDFOR	         
	\end{algorithmic} 
\end{algorithm}

\section{Prewhitening}
\label{sec:prew}
The fast NLS and proposed pitch tracking algorithm are derived under the assumption of white Gaussian noise. However, this assumption is usually violated in practice, for example, babble noise in a conference hall. Therefore, a prewhitening step is required to deal with the inconsistency between the white Gaussian noise model assumption and real life noise model. A linear prediction (LP) based prewhitening step is applied to each frame to deal with the non-white Gaussian noise (see \cite{norholm2016instantaneous,8683653} for detail). The power spectral density (PSD) of the noise given noisy signals is estimated using the minimum mean-square error (MMSE) estimator \cite{gerkmann2012unbiased}. We refer to the fast NLS and proposed algorithm with prewhitening step as Prew-Fast NLS and Prew-Proposed, respectively.

\section{Simulation}

\label{sec:sim} In this section, we test the performance of the proposed
harmonic model-based pitch tracking algorithm on real speech signals.

\subsection{Databases}
The databases used for evaluating the performance of different algorithms are as follows:

\emph{Keele pitch database}:  containing 10 spoken sentences from five male and five female speakers at a sampling rate of 20 kHz \cite{plante1995pitch}. the \textquotedbl{}ground truth\textquotedbl{} pitch estimates are extracted from electroglottography with 10 ms time frame increment and 25.6 ms frame length. In fact, there are many spikes and wrong estimates in the \textquotedbl{}ground truth\textquotedbl{} pitch values, especially in the transient frames. However, we present the results for the Keele database to facilitate comparison with other pitch estimation algorithms that use this database. 

\emph{Parkinson's disease database}: containing 130 sustained /a/ phonations from patients with Parkinson's disease \cite{tsanas2014robust} at a sampling rate of 44.1 kHz. Each of the phonations is in one second length. The estimated \textquotedbl{}ground truth\textquotedbl{} pitches in 10 ms time frame increment are extracted from electroglottography (EGG).

\subsection{Performance measures}
Three performance measures are considered:

\emph{Total error ratio (TER)} \cite{fisher2006generalized}: voicing detection performance measure. It is calculated based on the ratio between the
number of incorrect voicing detection (false positive and true negative) estimates 
and the number of total estimates. 

\emph{Gross error ratio (GER)} \cite{camacho2008sawtooth}: accuracy measure of pitch estimates. It is computed based on the
ratio between the number of pitch estimates that differ by more than 20 percents from
the ground truth and the number of total estimates. The unvoiced frames from the ground truth are excluded and the pitch value of the voiced frame that is wrongly labeled as unvoiced frames by different pitch estimation algorithms is set to 0. 

\emph{Mean absolute error (MAE)} \cite{tsanas2014robust}:  accuracy measure of pitch estimates. It is computed based on mean of the absolute errors between the ground truth and estimates.  The unvoiced frames from the ground truth are excluded and the oracle voicing detector is used for all the algorithms.

\subsection{Results}
%
%

\begin{figure}[t]
\includegraphics[scale=0.9]{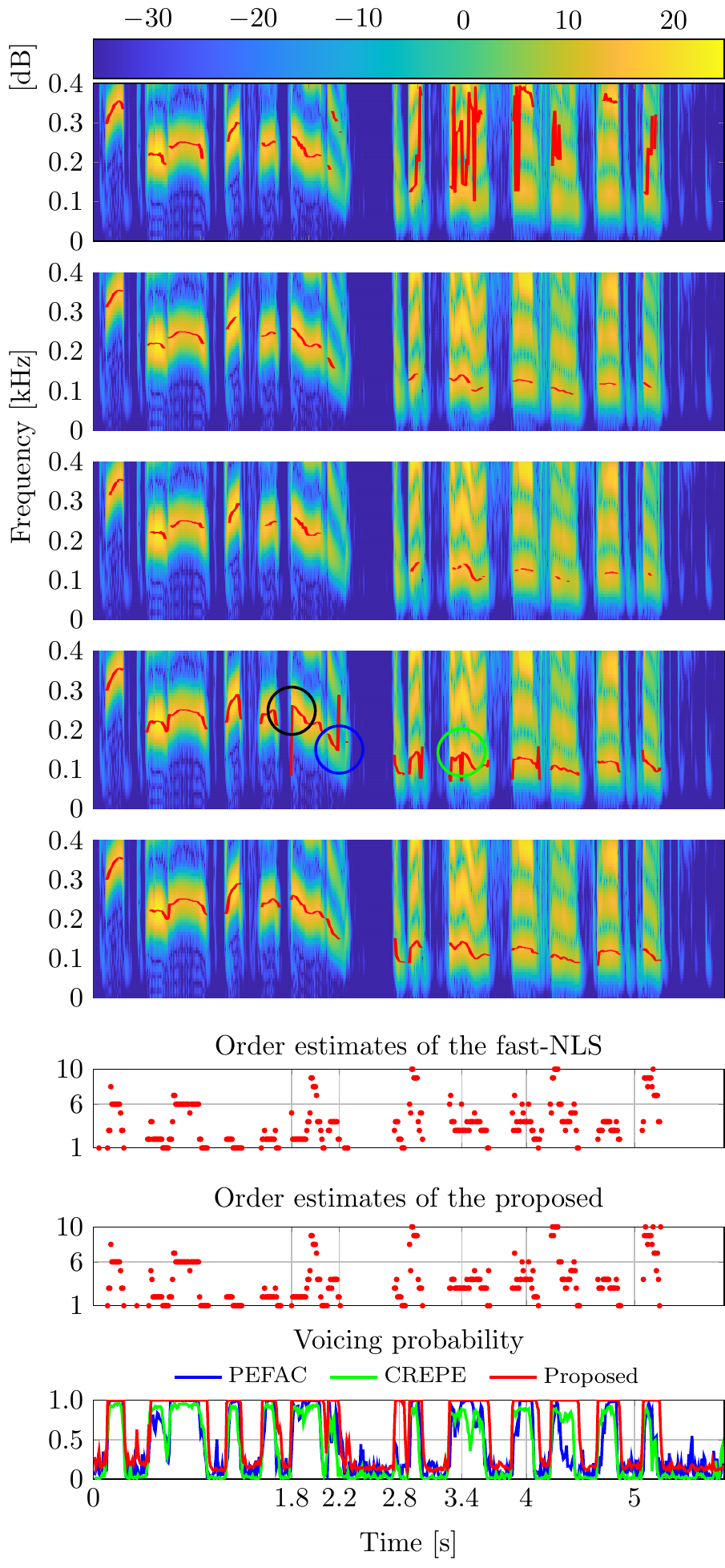}
\caption{Pitch estimates from PEFAC, CREPE, YIN, fast NLS and the proposed, the order estimates of the fast NLS and the proposed, and the voicing probabilities for real speech signals in 0 dB white Gaussian noise (from
top to bottom).}

\label{fig:toy} 
\end{figure}

First, the proposed approach is tested on concatenated speech signals
uttered by a female speaker first, male speaker second, sampled at
16 kHz. The spectrogram of the clean speech signals, pitch estimates,
order estimates and the voicing detection results for PEFAC, CREPE, YIN, fast NLS and
the proposed algorithm are shown in Fig.\ \ref{fig:toy}. The time
frames of the spectrograms without red lines on top are unvoiced or
silent frames. The variances for the transition matrices $\sigma_{\omega}^{2}$
and $\sigma_{K}^{2}$ are set to $\frac{16\pi^{2}}{f_{\text{s}}^2}$ and $1$,
respectively. The SNR for white Gaussian noise is set to 0 dB. The
candidate pitch $\omega_{0}$ is constrained to the range $2\pi\left[70\ 400\right]/f_{\text{s}}$ for PEFAC, YIN, fast NLS and the proposed algorithm. However, the results for the neural network based approach CREPE is based on the model with the pitch range $2\pi\left[32.7\ 1975.5\right]/f_{\text{s}}$ provided by the authors \cite{8461329}. To change the settings for CREPE, re-training of the neural network model is required. The maximum allowed harmonic
order for the proposed and fast NLS is set to 10. The frame length is $M=400$ samples (25 ms) with
$60\%$ overlap. As can be seen from Fig.\ \ref{fig:toy}, the voicing detection
results of both the fast NLS and the proposed algorithm are better
than those of YIN, PEFAC and CREPE. For example, the frames around 2.8 s are correctly classified as voiced by the fast NLS and the proposed, but wrongly labeled as unvoiced by YIN, PEFAC and CREPE. Fast NLS suffers from octave
errors, and has outliers particularly in the transition frames where voicing decisions change. In the transition frame around 1.8 s, the pitch and number of harmonics are wrongly estimated to 84.8 Hz and five, respectively, by the fast NLS. In contrast, they are estimated to 248.8 Hz and one, respectively, by the proposed. Clearly, the estimates of the proposed fit better to the spectrogram than the estimates of the fast NLS. The reason for the robustness against transient frames of the proposed algorithm is that the pitch and harmonic order information of the latest voiced frame is used as the prior, i.e. \eqref{pastinfo}. The harmonic order of the frame in 2.2 s is estimated to two by both the fast NLS and the proposed. However, the pitch is wrongly estimated to 288.8 Hz by the fast NLS, but correctly estimated to 150.4 Hz by the proposed. By imposing temporal smoothness prior on the pitch using the Markov process model $p(\omega_{n}|\omega_{n-1},u_{n}=1,u_{n-1}=1)$, smoother estimates of the pitches are obtained. An octave error is produced by the fast NLS in the frame around 3.4 s. The pitch and harmonic order are estimated to 72 and six, respectively, by the fast NLS, but 143.2 and three, respectively, by the proposed. In fact, harmonic orders are estimated to three in the surrounding frames by both the fast NLS and the proposed. By using Bayesian tracking for the pitches and harmonic orders, smoother estimates of the pitches and harmonic orders are obtained. In conclusion, the proposed Bayesian pitch tracking
algorithm obtains smooth estimates of the pitch and harmonic orders, and good voicing detection results by exploiting the past information.
\begin{figure}[t]
\includegraphics[scale=0.9]{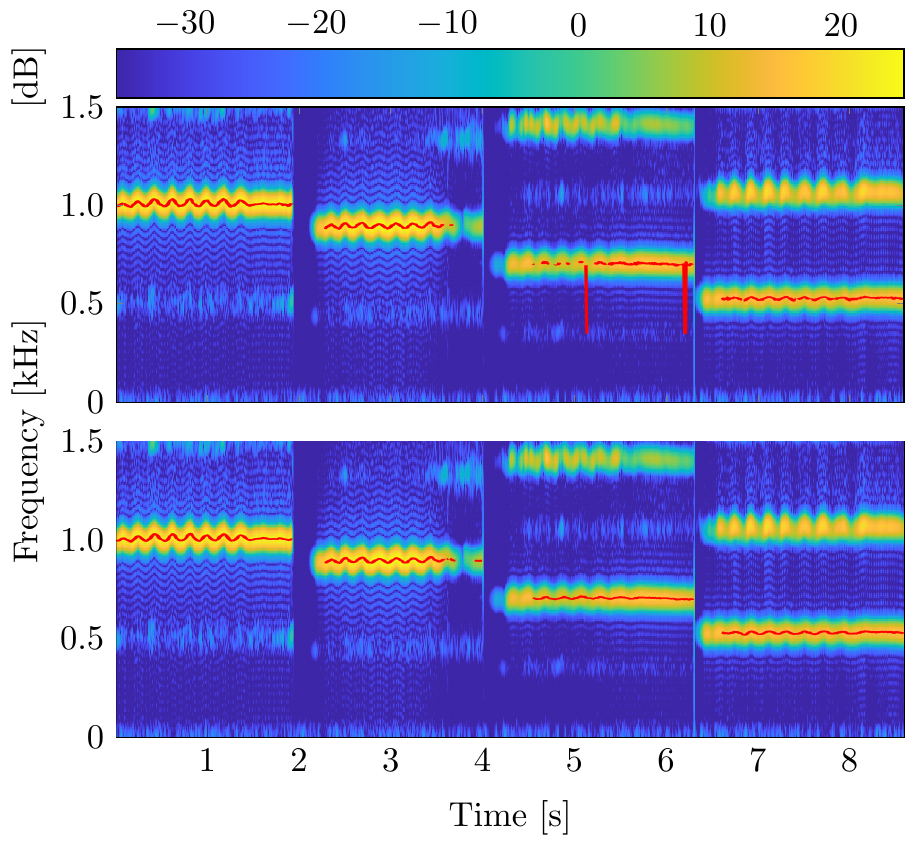}
\caption{Pitch estimates of fast NLS and the proposed algorithm for musical sounds in -5 dB white Gaussian noise (from top to
bottom).}
\label{fig:toymusic} 
\end{figure}

The second experiment tests the performance of the proposed algorithm
on musical instrument sounds (flute) from the University of Iowa Musical
Instrument Samples database, decreasing from note B5 to C5. The spectrogram
of the clean signals and the pitch estimates from fast NLS and the
proposed algorithm are shown in Fig.\ \ref{fig:toymusic}. The music
signal is downsampled to 16 kHz. The SNR for Gaussian white noise
is set to -5 dB. The pitch $\omega_{0}$ is constrained to the range
$2\pi\left[100\ 1500\right]/f_{\text{s}}$. The other parameters are set
to the same as for Fig.\ \ref{fig:toy}. As can be seen, the proposed
algorithm not only has smoother estimates of the pitch than fast NLS
but also better voicing detection results. 

\begin{figure}[t]
\includegraphics[scale=0.9]{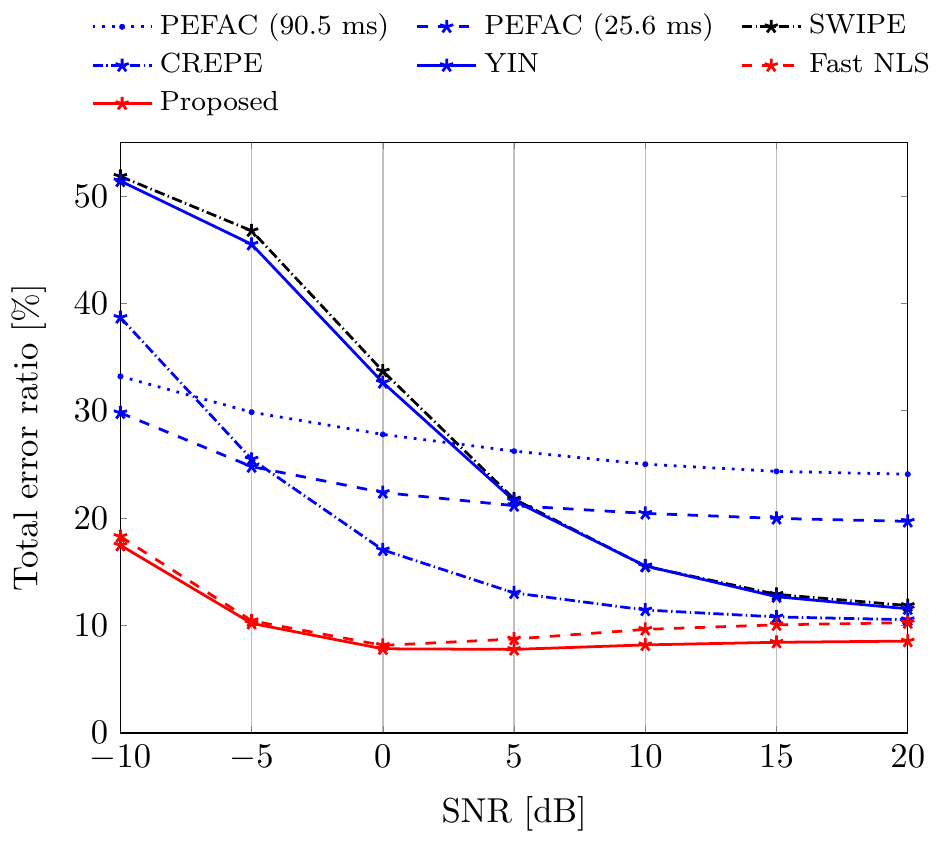}
\caption{Total error ratio in different SNRs for the Keele pitch database}
 \label{fig:ter} 
\end{figure}


\begin{figure}[t]
\includegraphics[scale=0.9]{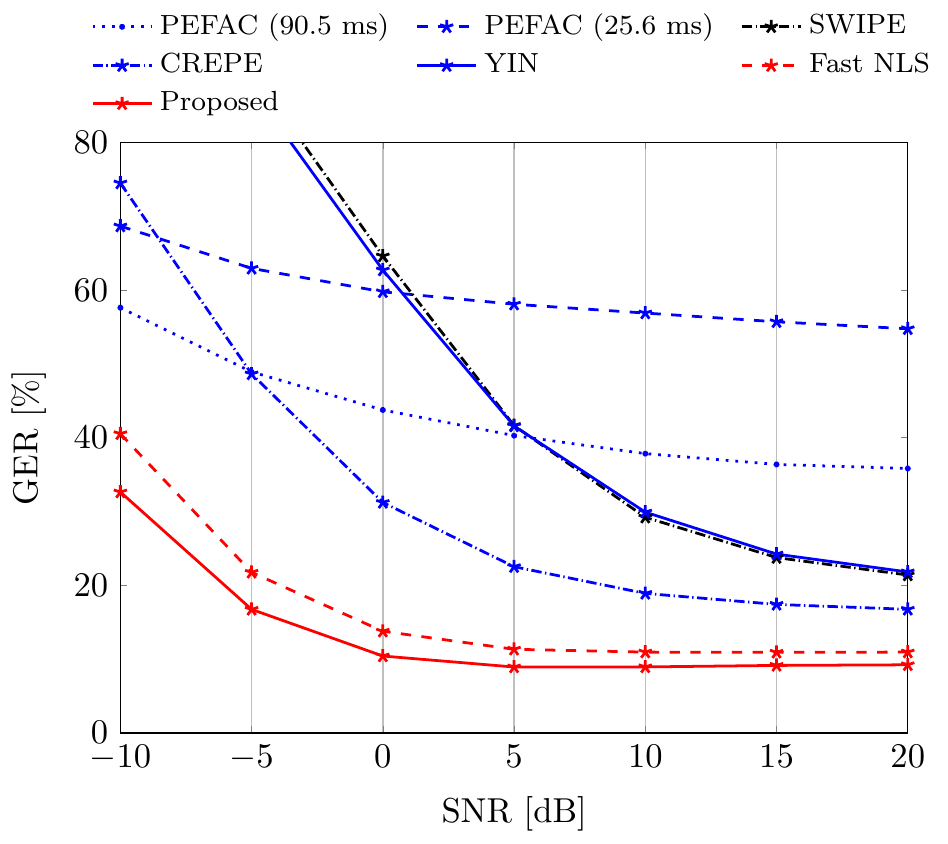}
\caption{Gross error ratio in different SNRs for the Keele pitch database}
 \label{fig:ger} 
\end{figure}

\begin{figure}[t]
\includegraphics[scale=0.9]{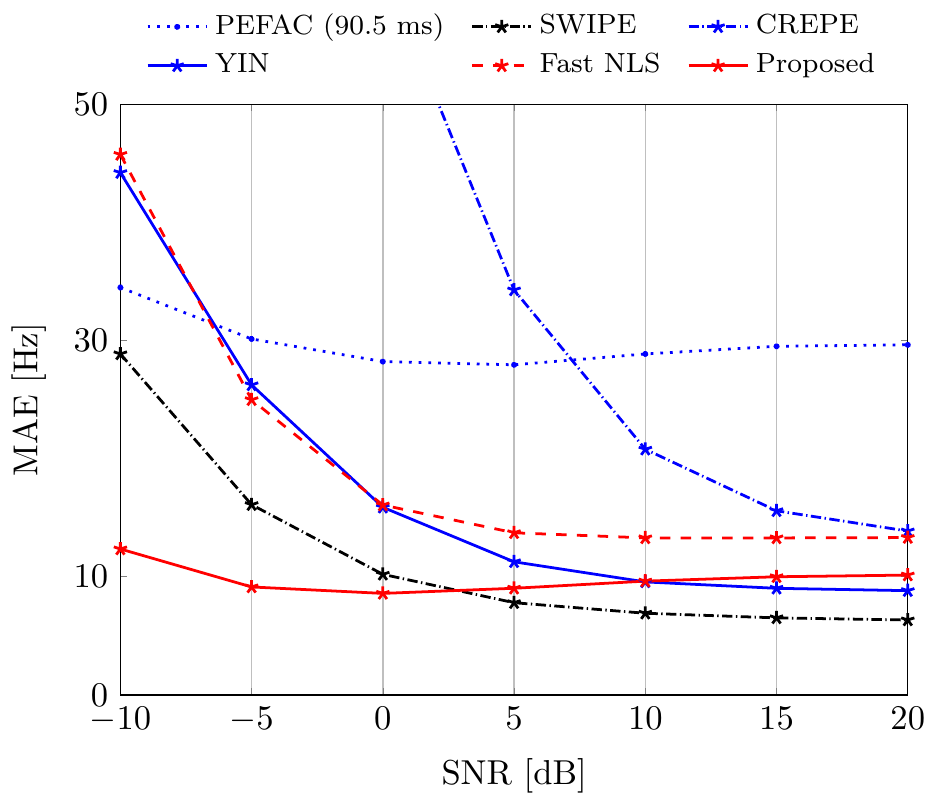}
\caption{Mean absolute error in different SNRs for the Keele pitch database with oracle voicing detector}
 \label{fig:mae} 
\end{figure}

\begin{table}
\caption{Total error ratio in colored noise}
\centering
\begin{tabular}{cccccc}
\toprule[.5pt]
\toprule[.5pt]
SNR	 	& &-5.00	 &0.00	 &5.00	 &10.00	 \\	\midrule[.5pt]
\multirow{2}{*}{PEFAC (90.5 ms)}	&Babble	 &0.42	 &0.38	 &0.34	 &0.29		 \\&Factory	&0.34	 &0.30	 &0.27	 &0.25	 \\ \midrule[.5pt]
\multirow{2}{*}{PEFAC (25.6 ms)}	&Babble	  &0.41	 &0.35	 &0.29	 &0.24	 \\&Factory	 &0.30	 &0.25	 &0.22	 &0.21	 \\ \midrule[.5pt]
\multirow{2}{*}{SWIPE}	&Babble	&0.50	 &0.42	 &0.29	 &0.19	 \\ &Factory	 &0.52	 &0.49	 &0.40	 &0.28	 \\  \midrule[.5pt]
\multirow{2}{*}{CREPE}	&Babble	 &0.40	 &0.29	 &0.21	 &0.16			 \\&Factory	 &0.39	 &0.28	 &0.20	 &0.15	 \\ \midrule[.5pt]
\multirow{2}{*}{YIN}	&Babble	 &0.50	 &0.43	 &0.32	 &0.22			 \\&Factory	 &0.50	 &0.43	 &0.32	 &0.22	 \\ \midrule[.5pt]
\multirow{2}{*}{Prew-Fast NLS}	&Babble	 &0.35	 &0.27	 &0.18	 &0.12	 \\&Factory	&0.28	 &0.20	 &\textbf{0.14}	 &\textbf{0.11}	 \\ \midrule[.5pt]
\multirow{2}{*}{Prew-Proposed}  &Babble	  &\textbf{0.34}	 &\textbf{0.25}	 &\textbf{0.17}	 &\textbf{0.12} 	\\&Factory	 &\textbf{0.28}	 &\textbf{0.20}	 &0.15	 &0.12	 \\ 
\bottomrule[.5pt]
\bottomrule[.5pt]
\end{tabular}
\label{table:MyLabelTER}
\end{table}


\begin{table}
\caption{Gross error ratio in colored noise}
\centering
\begin{tabular}{cccccc}
\toprule[.5pt]
\toprule[.5pt]
SNR	&   &-5.00	 &0.00	 &5.00	 &10.00	 \\	\midrule[.5pt]
\multirow{2}{*}{PEFAC (90.5 ms)}	&Babble  &0.62	 &0.51	 &0.44	 &0.39		 \\  &Factory &0.56	 &0.47	 &0.41	 &0.38	 \\ \midrule[.5pt]
\multirow{2}{*}{PEFAC (25.6 ms)}	&Babble  &0.72	 &0.65	 &0.60	 &0.57	 \\ &Factory &0.68	 &0.61	 &0.57	 &0.54	 \\ \midrule[.5pt]
\multirow{2}{*}{SWIPE}	&Babble  &0.96	 &0.81	 &0.55	 &0.36	 \\ &Factory &1.00	 &0.94	 &0.76	 &0.54	 \\ \midrule[.5pt]
\multirow{2}{*}{CREPE }	&Babble  &0.73	 &0.50	 &0.34	 &0.24	 \\  &Factory &0.75	 &0.53	 &0.36	 &0.26	 \\ \midrule[.5pt]
\multirow{2}{*}{YIN }	&Babble  &0.95	 &0.83	 &0.61	 &0.42	 \\  &Factory &0.96	 &0.83	 &0.61	 &0.42	 \\ \midrule[.5pt]
\multirow{2}{*}{Prew-Fast NLS}&Babble 	 &0.57	 &0.41	 &0.30	 &0.24	 \\ &Factory &0.55	 &0.42	 &0.33	 &0.28	 \\ \midrule[.5pt]
\multirow{2}{*}{Prew-Proposed}	&Babble &\textbf{0.53}	 &\textbf{0.36}	 &\textbf{0.27}	 &\textbf{0.24}	 \\ &Factory &\textbf{0.51}	 &\textbf{0.37}	 &\textbf{0.29}	 &\textbf{0.25}	 \\ 
\bottomrule[.5pt]
\bottomrule[.5pt]
\end{tabular}
\label{table:MyLabelGER}
\end{table}

%

\begin{table}
\caption{Mean absolute value [Hz] in colored noise with oracle voicing detector}
\centering
\begin{tabular}{cccccc}
\toprule[.5pt]
\toprule[.5pt]
SNR	  &    &-5.00	&0.00	 &5.00	 &10.00	 \\	\midrule[.5pt]
\multirow{2}{*}{PEFAC (90.5 ms)}&Babble 	 &49.81	 &39.15	 &31.73	 &27.96	 \\&Factory&36.20	 &31.24	 &27.97	 &26.69	 \\ \midrule[.5pt]
\multirow{2}{*}{PEFAC (25.6 ms)}&Babble 	  &81.49	 &72.65	 &65.71	 &60.54	 \\&Factory &72.61	 &64.93	 &57.93	 &54.20		 \\ \midrule[.5pt]
\multirow{2}{*}{SWIPE}	&Babble &\textbf{31.73}	 &\textbf{17.94}	 &\textbf{10.95}	 &\textbf{8.04}	 \\ &Factory &43.91	 &27.02	 &16.02	 &10.51 \\ \midrule[.5pt]
\multirow{2}{*}{CREPE}	&Babble  &68.95	 &44.93	 &30.57	 &21.89	 \\ &Factory &79.00	 &52.41	 &34.51	 &24.70	 \\ \midrule[.5pt] 
\multirow{2}{*}{YIN}	&Babble  &56.25	 &39.05	 &23.86	 &14.96	 \\ &Factory &57.37	 &38.53	 &23.41	 &14.97	 \\ \midrule[.5pt]
\multirow{2}{*}{Prew-Fast NLS}	 &Babble  &64.81	 &45.79	 &31.45	 &23.79	 \\&Factory  &74.58	 &57.88	 &44.93	 &36.50	 \\ \midrule[.5pt]
\multirow{2}{*}{Prew-Proposed}	&Babble &33.33	 &17.91	 &12.22	 &10.81	 \\ &Factory   &\textbf{19.32}	 &\textbf{13.20}	 &\textbf{11.23}	 &\textbf{10.48}	 \\ 
\bottomrule[.5pt]
\bottomrule[.5pt]
\end{tabular}

\label{table:MyLabelMAE}
\end{table}

%

In the third experiment, we test the performance of the proposed algorithm
on the Keele pitch database with white Gaussian noise. Averages over 20 independent Monte Carlo experiments are used to obtain the experimental results. TER,  GER and  MAE for different
SNRs for PEFAC, SWIPE, YIN, CREPE, fast NLS and the proposed algorithm are shown
in Fig.\ \ref{fig:ter}, Fig.\ \ref{fig:ger} and  Fig.\ \ref{fig:mae}, respectively. 
For YIN, fast NLS and the proposed algorithm, the frame length is
set to the same as the reference, i.e., 25.6 ms. Frame lengths 25.6
ms and 90.5 ms (default value) are used for PEFAC.  The other parameters are set
to the same as for Fig.\ \ref{fig:toy}.  As can be seen from Fig.\ \ref{fig:ter} and Fig.\ \ref{fig:ger},
PEFAC has better performance in terms of both GER and TER than CREPE
at -10 dB SNR. Moreover, using a longer frame length (90.5 ms) for PEFAC
leads to a lower GER but a higher TER compared with a shorter frame
length (25.6 ms). SWIPE and YIN have similar performance in terms of TER and GER. The fast NLS method outperforms the PEFAC, SWIPE, YIN and CREPE. By imposing a smoothing prior on the pitches, harmonic orders and the  voicing and using the harmonic model combined, the proposed algorithm achieves lower GER and TER than the fast-NLS. As can be seen from Fig.\ \ref{fig:mae}, when the oracle voicing detector is used, the SWIPE has the lowest MAE from  5 to 20 dB while the proposed algorithm achieves the best performance from -10 to 0 dB.

In the fourth experiment, the performance of the proposed algorithm with prewhitening is tested on the Keele pitch database in colored noise, i.e., babble noise and factory noise. The TER, GER and MAE results for Prew-proposed, Prew-fast NLS, PEFAC, Yin and SWIPE are shown in \ref{table:MyLabelTER}, \ref{table:MyLabelGER} and \ref{table:MyLabelMAE}, respectively. The linear prediction order for the prewhitening is set to 30. The maximum allowed harmonic order for the proposed and fast NLS is set to 30. The other parameters are set
to the same as for Fig.\ \ref{fig:ter}. As can be seen from TABLE \ref{table:MyLabelTER} and \ref{table:MyLabelGER}, PEFAC with 90.5 ms and 25.6 ms have a lower TER and GER than YIN and SWIPE in -5 and 0 SNR conditions. The Prew-Proposed and Prew-Fast NLS have lower voicing detection errors and Gross errors than YIN, PEFAC and SWIPE in both babble and factory noise conditions.  Although similar performance in term of TER can be seen for Prew-Proposed and Prew-Fast NLS, the Prew-Proposed has a lower GER than Prew-Fast NLS. As can be seen from TABLE \ref{table:MyLabelMAE}, when the oracle voicing detector is used, the SWIPE achieves the lowest MAE in babble noise. The Prew-proposed has a comparable performance with the SWIPE in babble noise and has the best performance in factory noise.

\begin{figure}[t]
\includegraphics[scale=0.9]{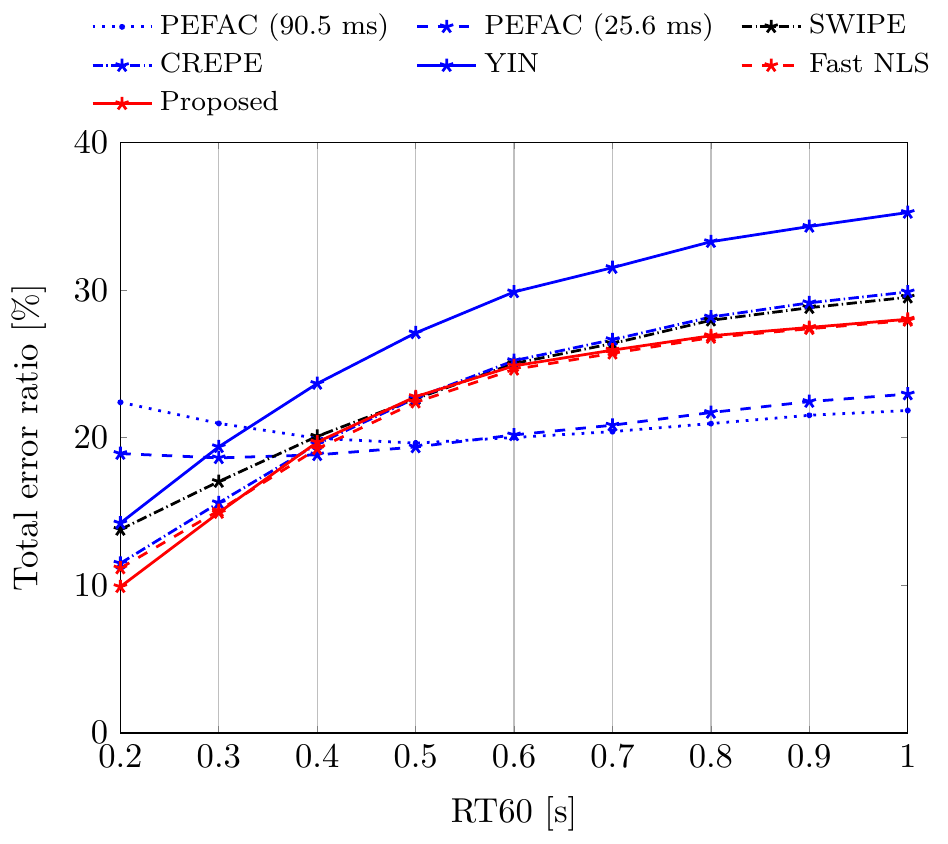}
\caption{Total error ratio in different reverberation time for the Keele pitch database}

 \label{fig:ter2} 
\end{figure}

\begin{figure}[t]
\includegraphics[scale=0.9]{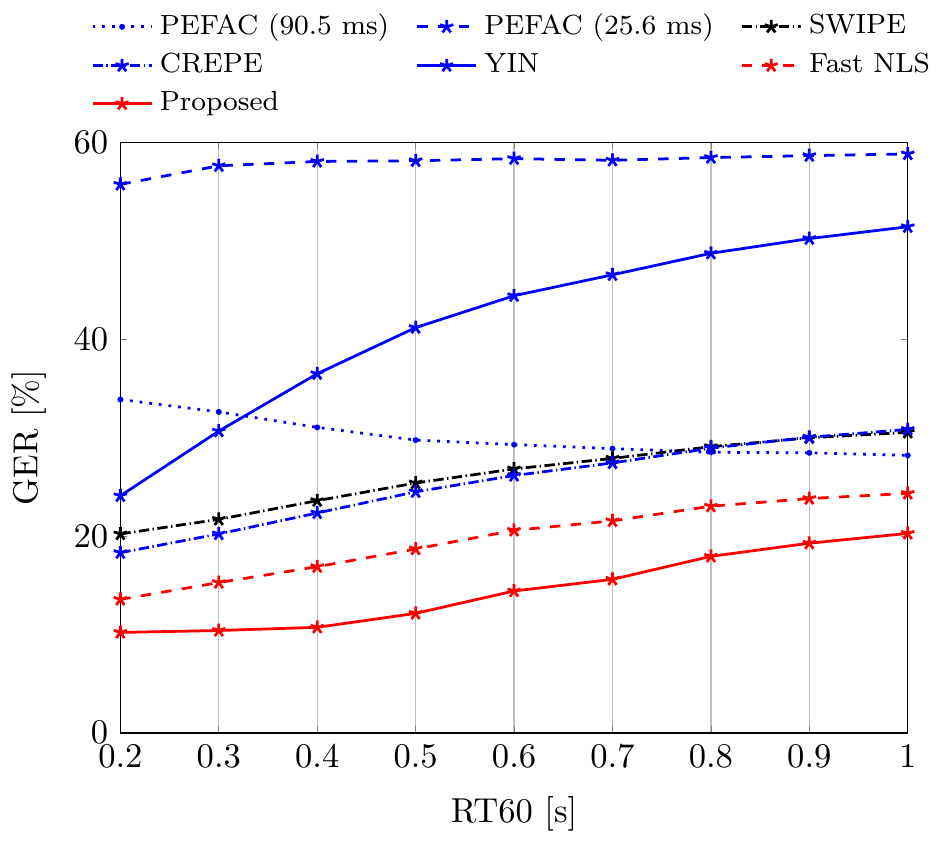}
\caption{Gross error ratio in different reverberation time for the Keele pitch database}

 \label{fig:ger2} 
\end{figure}

\begin{figure}[t]
\includegraphics[scale=0.9]{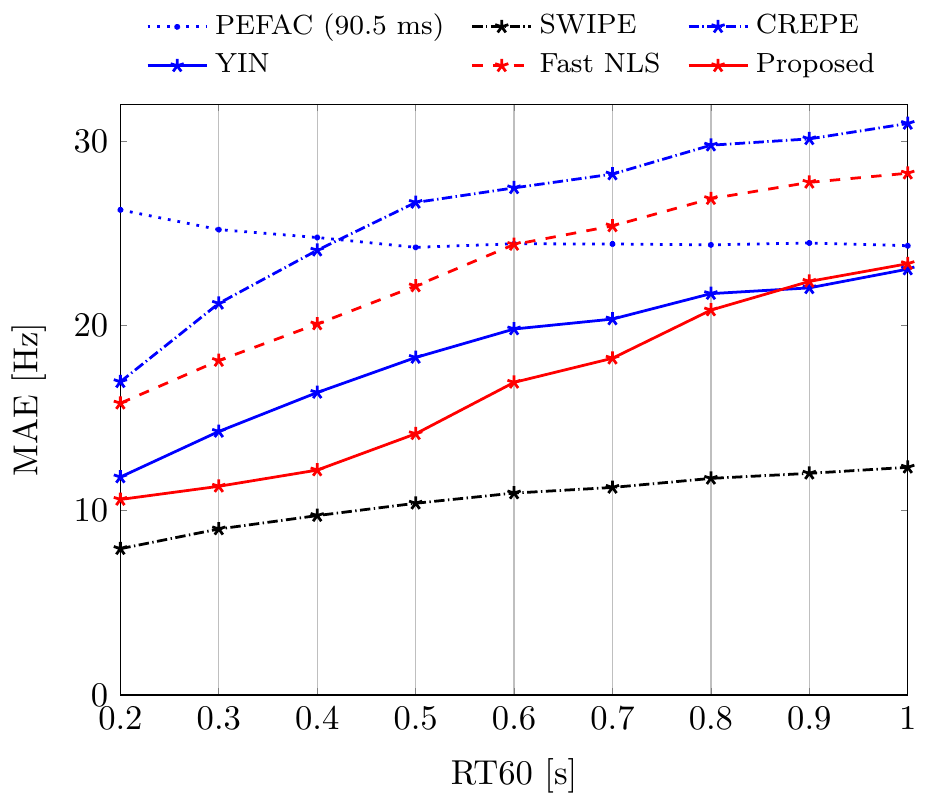}
\caption{Mean absolute error in different reverberation time for the Keele pitch database  with oracle voicing detector}

 \label{fig:mae2} 
\end{figure}

In the fifth experiment, we investigate the effect of reverberation on the performance of different pitch estimation algorithms. Reverberation is the process of multi-path propagation and occurs when the speech or audio signals are recorded in an acoustically enclosed space.  A commonly used metric to measure the reverberation is the reverberation time  (RT60) \cite{naylor2010speech}. The reverberated signals used for testing are generated by filtering the signal by synthetic room impulse responses (RIRs) with RT60 varying from 0.2 to 1 s in 0.1 s step. The dimension of the room is set to $10 \times 6 \times 4$ m. The distance between the source and microphone is set to 1 m. The  RIRs are generated using the image method \cite{allen1979image} and implemented using the RIR Generator toolbox \cite{habets2006room}. The position of the receiver is fixed while the position of the source is varied randomly from 60 degrees left of the receiver to 60 degrees right of the receiver for each Monte Carlo experiment. The TER, GER and MAE results on the Keele pitch database for the proposed, fast NLS, PEFAC, Yin and SWIPE are shown in Fig.\ \ref{fig:ter2}, Fig.\ \ref{fig:ger2} and Fig.\ \ref{fig:mae2}, respectively, where the parameters are set
to the same as for Fig.\ \ref{fig:ter}. As can be seen from Fig.\ \ref{fig:ter2}, the PEFAC (90.5 ms) has the lowest voicing detection errors in more reverberated conditions (RT60 from 0.5 to 1 s) while the proposed algorithm has a better voicing detection performance in less reverberated conditions. The proposed and the  fast NLS has similar performance in terms of TER. However, as can be seen from Fig.\ \ref{fig:ger2}, the proposed outperforms the PEFAC, SWIPE, CREPE, YIN and fast NLS in terms of GER. From Fig.\ \ref{fig:mae2}, we can conclude that SWIPE has the best performance while the proposed is the second best one  in terms of MAE.

%
%

\begin{figure}[t]
\includegraphics[scale=0.9]{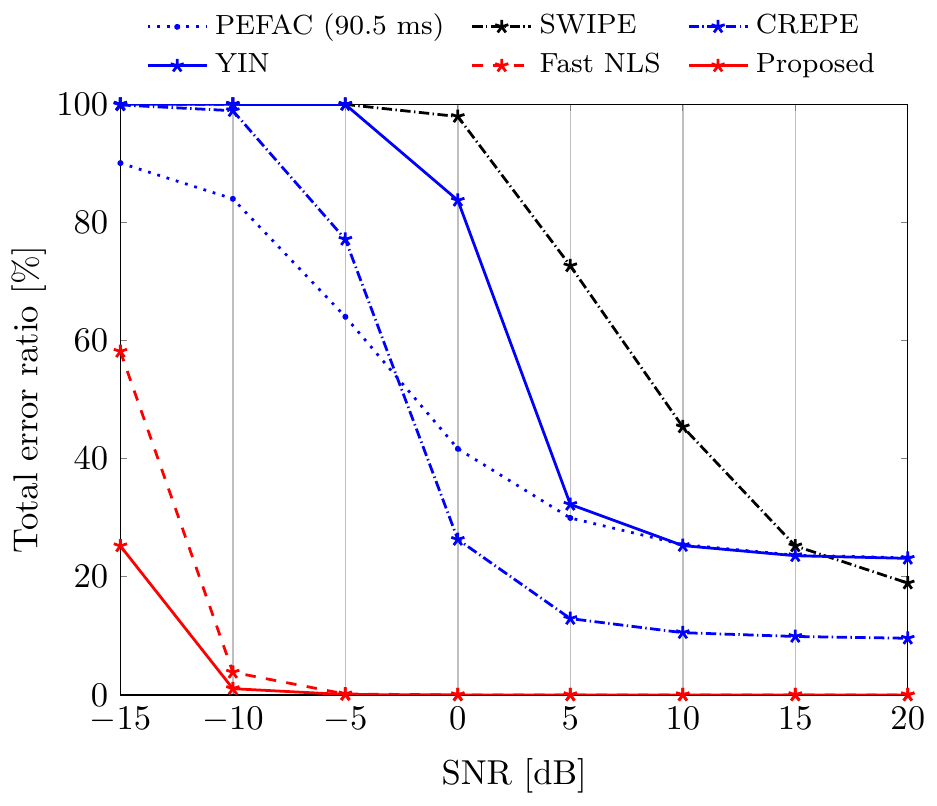}
\caption{Total error ratio in different SNRs for the Parkinson's disease database}

 \label{fig:ter1} 
\end{figure}

\begin{figure}[t]
\includegraphics[scale=0.9]{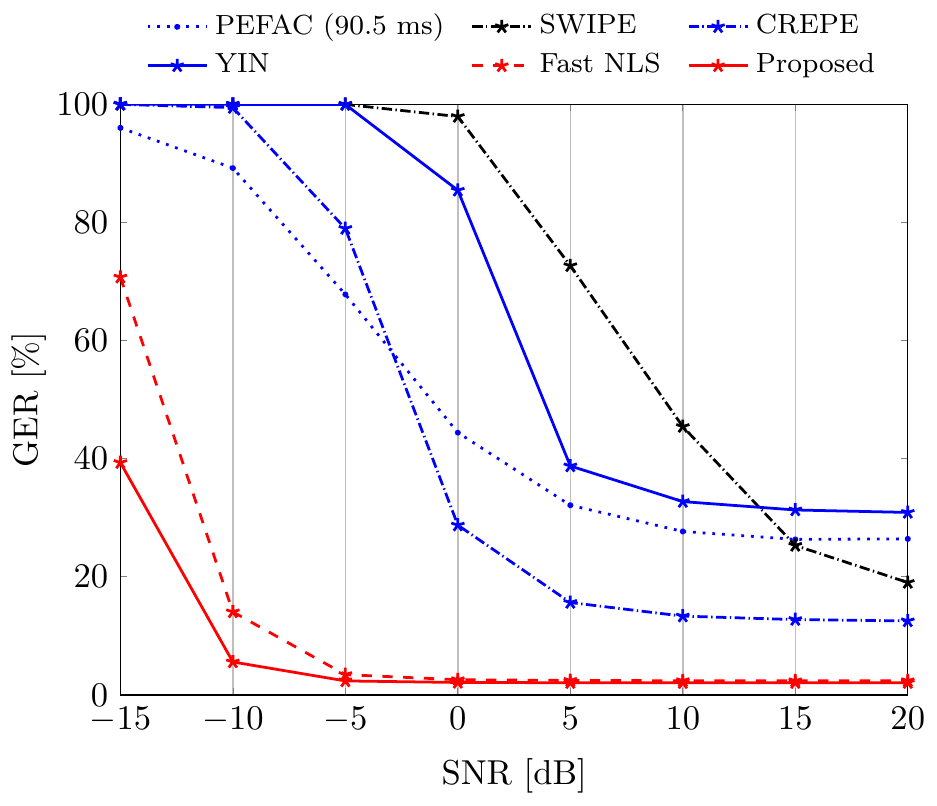}
\caption{Gross error ratio in different SNRs for the Parkinson's disease database}

 \label{fig:ger1} 
\end{figure}

\begin{figure}[t]
\includegraphics[scale=0.9]{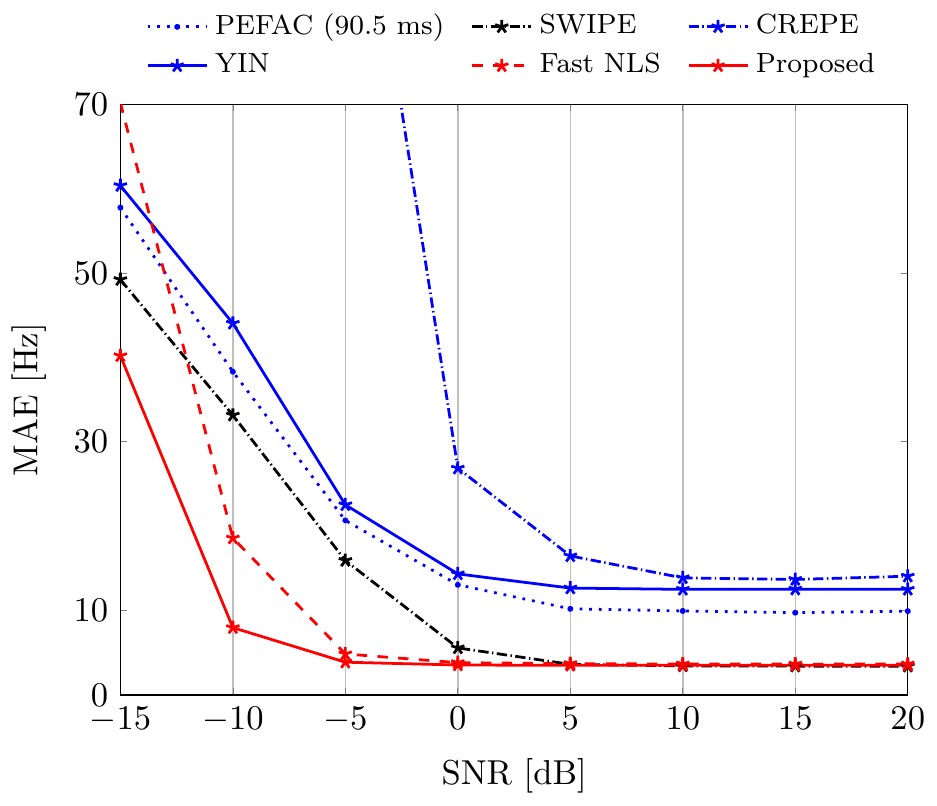}
\caption{Mean absolute error in different SNRs for the Parkinson's disease database with oracle voicing detector}

 \label{fig:mae1} 
\end{figure}

\begin{figure}[t]
\includegraphics[scale=0.9]{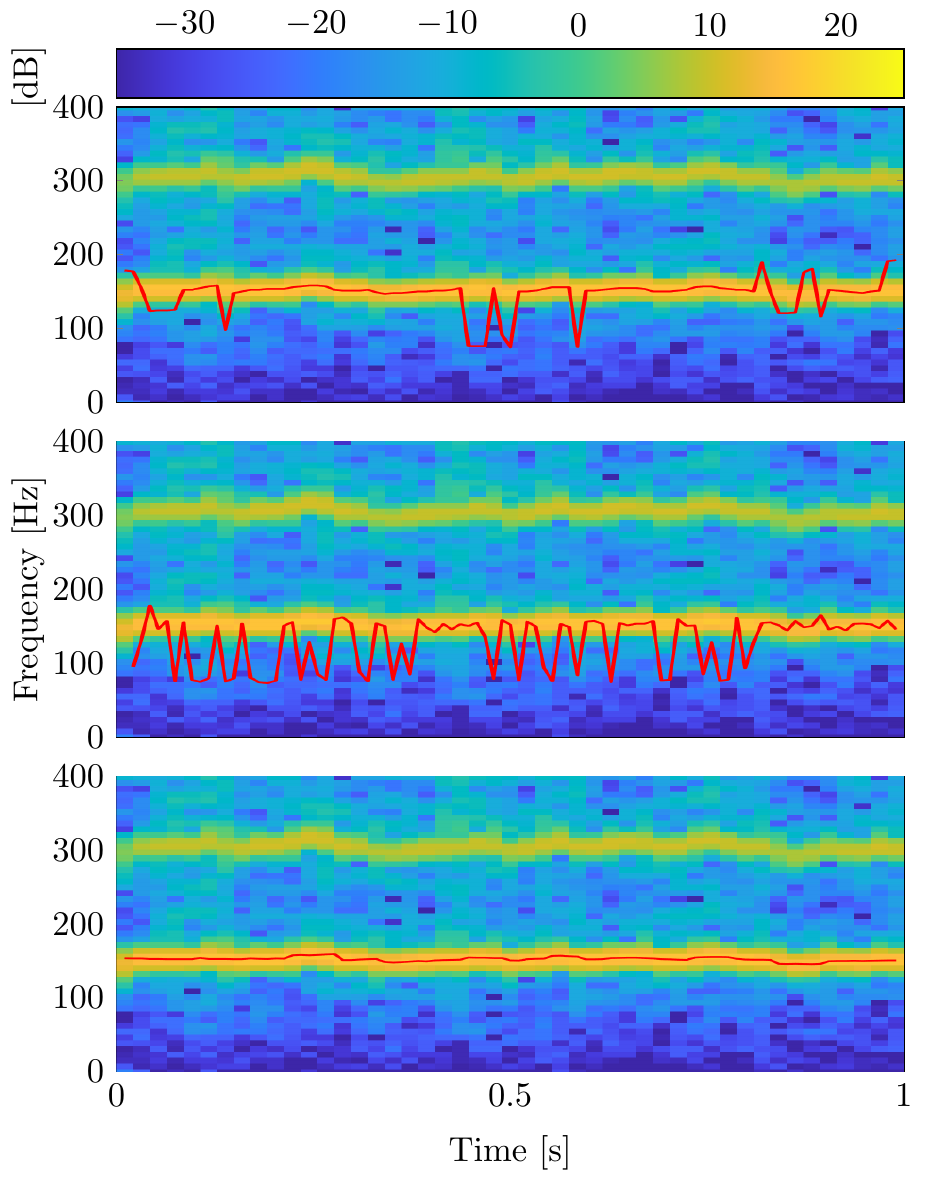}
\caption{Pitch estimates from PEFAC (oracle), YIN (oracle), and the proposed for sustained /a/ sounds from a database of Parkinson's disease voices in 0 dB white Gaussian noise.}
\label{fig:toya}
\end{figure}

In the final experiment, the performance of the proposed algorithm
is tested on sustained /a/ signals (voiced) from the Parkinson's disease database. The signals are downsampled to 16 kHz. TER, GER and MAE for different SNRs are shown
in Fig.\ \ref{fig:ter1}, Fig.\ \ref{fig:ger1} and Fig.\ \ref{fig:mae1}, respectively.  The parameters are set
to the same as for Fig.\ \ref{fig:ter}.  Similar conclusions
to Fig.\ \ref{fig:ter} and Fig.\ \ref{fig:ger} can be drawn from
Fig.\ \ref{fig:ter1} and Fig.\ \ref{fig:ger1}. The proposed algorithm
has the best performance in terms of the TER and GER. Moreover, the proposed algorithm presents the lowest MAE, especially from -15 to 0 dB. The spectrogram of one of the sustained /a/ sounds from the Parkinson's disease database, pitch estimates of the PEFAC (oracle), YIN (oracle) and the proposed algorithm in 0 dB white Gaussian noise are shown in Fig.\ \ref{fig:toya}. The oracle voicing detector from the ground truth (all voiced) is used for both PEFAC and YIN. As can be seen from Fig.\ \ref{fig:toya}, the proposed algorithm outperforms the PEFAC (oracle) and YIN (oracle).

Based on the above experiments, PEFAC obtains a better pitch estimation and voicing detection performance than the neural network-based CREPE in low SNR scenarios. The SWIPE offers good performance in terms of MAE in high SNRs. The proposed algorithm obtains superior performance in terms of GER, TER and MAE compared to PEFAC, SWIPE,YIN, CREPE, and the fast NLS in low SNR scenarios for the Keele pitch database and Parkinson's disease database.

\section{Conclusions}

\label{sec:con} In this paper, a fully Bayesian harmonic model-based
pitch tracking algorithm is proposed. Using a parametric harmonic
model, the proposed algorithm shows good robustness against noise.
The non-stationary evolution of the pitch, harmonic order and voicing
state are modelled using first-order Markov chains. A fully Bayesian approach is applied for the noise variance and weights to avoid over-fitting. Using the hierarchical g-prior for the weights, the likelihood
function can be easily evaluated using the fast NLS. The computational
complexity of the recursive calculation of the predicted and posterior
distributions is reduced by exploiting conditional independence between
the pitch and harmonic order given the voicing indicators. Simulation
results show that the proposed algorithm has good robustness against
voicing state changes by carrying past information on pitch over the
unvoiced/silent segments. The results of the pitch estimates and voicing detection
for spoken sentences and sustained vowels are compared against ground
truth estimates in the Keele and Parkinson's disease databases, showing
that the proposed algorithm presents good pitch estimation and voicing detection accuracy even in very noisy conditions (e.g., -15 dB).

\ifCLASSOPTIONcaptionsoff \newpage{}\fi

\end{document}